\documentclass{aa}

\usepackage{hyperref}
\usepackage{amsmath}
\usepackage{bm}
\usepackage{subcaption}
\usepackage{booktabs}
\usepackage{graphicx}
\usepackage{txfonts}
\usepackage{lipsum}
\usepackage{subcaption}
\usepackage{lscape}
\usepackage{placeins}

\begin{document}

\title{GPU-accelerated X-ray pulse profile modeling}

\author{Tianzhe Zhou\inst{1}\fnmsep\thanks{ zhoutz22@mails.tsinghua.edu.cn}
\and Chun Huang\inst{2}
}

\institute{Department of Physics, Tsinghua University, Beijing 100084, China
\and Physics Department and McDonnell Center for the Space Sciences, Washington University in St. Louis; MO, 63130, USA}

\date{Received Febrary 3, 2026}

\abstract
{}
{The aim of this work is to remove the accuracy-speed bottleneck in Bayesian pulse profile modeling (PPM) of thermal X-ray emission from rotation-powered millisecond pulsars. This would enable a reliable inference of the stellar mass, $M$, radius, $R$, and the equation of state of cold, dense matter for extreme hotspot geometries and higher fidelity forward models.}
{We developed, validated, and publicly released a GPU-accelerated X-ray PPM framework. We benchmarked the GPU implementation against established reference calculations across standard and extreme geometries. We quantified the performance on an RTX~4080. We also diagnosed numerical systematics associated with atmosphere-table interpolation near lookup boundaries using two targeted tests and mitigated them with a mixed-order interpolation scheme.}
{Our framework reproduces established benchmarks to a $\sim10^{-3}$ relative accuracy, including the case of extreme hotspot configurations that are difficult to resolve at production settings. At high fidelity, we reduced per-evaluation runtimes from minutes to $2$--$5$~ms on an RTX~4080, corresponding to $10^{3}$--$10^{4}\times$ speedups, thereby making posterior exploration feasible at resolutions and model complexities that were previously impractical. We identified a systematic error near atmosphere-table interpolation boundaries and show that the proposed mixed-order interpolator substantially reduces this bias in diagnostic tests.}
{By coupling benchmark-level accuracy with millisecond-scale evaluations, the framework expands the accessible hotspot model space and mitigates key numerical systematics in X-ray PPM, strengthening mass-radius inference for current and future X-ray missions.}


\maketitle

\nolinenumbers

\section{Introduction} \label{sec:intro}
Rotation-induced modulations of emission from neutron-star
surfaces known as X-ray pulse profiles  encode the surface temperature distribution and the global parameters: mass, $M$, and radius, $R$. The emission is often dominated by one or more localized, hot regions (i.e., hotspots). Photons from these regions experience gravitational redshift and strong-field light bending, as well as special-relativistic Doppler boosting and time delays. These effects leave signatures in the energy-resolved waveform, encoding the star's compactness, our primary interest, as well as other global and geometric parameters \citep{Pechenick_1987,Beloborodov_2002,Poutanen_2006,Morsink_2007,Cadeau_2007,Psaltis_2014,Watts_2016,Nattila_2018,Poutanen_2020}. Although this picture is conceptually clean, to model observed pulse profiles accurately, and to extract the hotspot configuration and mass-radius ($M$-$R$) relation, we must incorporate the realistic atmospheric beaming pattern, including composition and magnetic-field effects \citep{zavlin_96,Ho_2001,Ho_2003,Suleimanov_2012,Salmi_2020,Salmi_2023}, rotation-induced oblateness \citep{Miller1998,Poutanen_2003,Morsink_2007,AlGendy_14,Nattila_2018,jakab2025gravitationalredshiftrapidlyrotating}, hotspot geometry \citep{Poutanen_2006,Viironen_04,Bogdanov_2008}, interstellar absorption \citep{verner_1996,wilms_00}, and the detailed instrument response \citep{Gendreau2016,Prigozhin_16,Remillard_2022,Markwardt_24}. With these ingredients and practical approximations established, reliable forward modeling procedures can be carried out across the full parameter space, enabling inference of $M$ and $R$ using the X-ray pulse profile modeling (PPM) technique \citep{Bogdanov_19,Bogdanov_21,Miller2019,Riley2019,Miller2021,Riley2021,Salmi_2024_j0740,Choudhury24,Vinciguerra24,Dittmann24}.

Thermal emission from neutron-star surfaces peaks in the soft X-ray band ($\sim 0.1$--$2 \,\mathrm{keV}$) \citep{zavlin_96,Potekhin_2015} and, thus, obtaining precise pulse profile measurements requires high throughput at these energies, sub-microsecond time tagging, and good energy resolution. NASA’s {Neutron Star Interior Composition Explorer} (NICER) provides this capability: a $0.2$--$12\,\mathrm{keV}$ bandpass, large effective area at soft X-rays, $\sim 100\,\mathrm{ns}$ absolute timing, and $\sim 10^2\,\mathrm{eV}$ energy resolution near $1\,\mathrm{keV}$ \citep{Gendreau2016,Prigozhin_16}. NICER’s primary pulse profile targets are rotation-powered millisecond pulsars (MSPs), whose highly coherent spin permits phase-folding of long observations to build  energy-resolved waveforms with high signal-to-noise
ratio. Statistical uncertainties then decrease roughly as the square root of the total photon counts (or exposure time). Accordingly, NICER has concentrated on a small set of bright, well-characterized MSPs that would be suitable for running the PPM \citep{Bogdanov_19_dataset,Guillot_19_dataset}. Looking ahead, the upcoming {enhanced X-ray Timing and Polarimetry mission} ({eXTP}), currently planned for launch in the early 2030s, is designed to combine large-area timing, spectroscopy, and polarimetry, and is expected to make PPM an even more powerful probe of neutron-star dense matter \citep{eXTP_Zhang,eXTP_Li,eXTP_Watts}.

The first landmark NICER PPM result came from PSR~J0030+0451, an {isolated} millisecond pulsar with no prior radio mass measurement \citep{Manchester_2005,Lommen_2006}. Independent analyses of the same NICER dataset produced joint posteriors for the stellar mass and radius and favored nonantipodal emitting-region geometries, establishing the first NICER mass-radius constraints via PPM. Reanalyses have refined these constraints \citep{Miller2019,Riley2019,Vinciguerra24}. NICER’s subsequent measurement for PSR~J0740+6620, whose mass is tightly constrained by radio timing at $\simeq 2\,M_\odot$ \citep{Cromrtie_20,Fonseca_21}, yielded radius measurements at the high-mass end from independent analyses \citep{Miller2021,Riley2021}, with subsequent updates using additional NICER data \citep{Salmi_2024_j0740,Dittmann24} placing strong constraints on the dense-matter equation of state and disfavoring very soft models that predict markedly smaller radii (e.g., \citealt{Raaijmakers2021,Zhang_21}). In both sources, the inferred hotspot locations are inconsistent with a simple centered dipole, suggesting more complex (e.g., offset or multipolar) surface-field configurations \citep{Bilous_2019,Petri:2025nhu,Kalapotharakos_2021,Chen_2020,Huang:2025_hotspot}. Recent and ongoing NICER campaigns have targeted bright MSPs such as PSR~J0437$-$4715, whose radio timing yields a precise mass of $M=1.418\pm0.044\,M_\odot$, and PSR~J0614$-$3329, thereby extending radius measurements toward the canonical $\sim 1.4\,M_\odot$ mass scale \citep{Reardon_16,Reardon_2024,Choudhury24,mauviard2025nicerview14solarmass}.

Together with tidal-deformability constraints from gravitational-wave events \citep{Abbott_2017,Abbott_2018,Abbott_2019}, these results have energized the nuclear-astrophysics community and motivated increasingly parametric and nonparametric equation-of-state inference frameworks that confront neutron-star structure with multimessenger data (e.g., \citealt{Soumi_2018_GW_PRL,landry_2019,Raaijmakers_2019,Raaijmakers_2020,Raaijmakers_2021,Essick_2020,Greif_2019,Legred_2021,Huth_2022,Annala_2022,Huang:2023grj,Huang:2024rfg,Huang:2024rvj,Rutherford24,Zhou:2025uim,Cartaxo:2025jpi,Huang:2025vfl,Huang:2024wig,Huang:2025mrd}).
Since NICER’s initial PPM results, the applications and methodological advancements of PPM have expanded substantially. Open-source frameworks, such as \texttt{X-PSI}, have made fully Bayesian PPM accessible and reproducible, providing flexible hotspot parameterizations \citep{XPSI_2021,XPSI_joss}. 

Reanalyses of PSR~J0030+0451 have shown that the choice of hotspot maps and the treatment of instrumental and background systematics can shift the inferred $M$--$R$ posteriors, especially when joint NICER+XMM constraints are imposed \citep{Vinciguerra24}. This approach has since been extended to additional NICER targets, notably PSR~J0437$-$4715 \citep{Choudhury24}, and recent studies of PSR~J1231$-$1411 \citep{Salmi_2024_Complex}. In parallel, key physical ingredients continue to be refined and several formalisms now propagate polarization predictions suitable for cross-mission tests \citep{Heyl_2000,Viironen_04,Loktev_2020,Salmi_2025}. Methodological progress also includes faster and more accurate relativistic ray-tracing schemes \citep{Nattila_2018,Poutanen_2020}. Overall, PPM remains a rapidly advancing field.

Despite major advances, several bottlenecks still limit current PPM analyses. First, degeneracies among hotspot geometry and stellar compactness can broaden posterior inferences \citep{Poutanen_2006,Lo_2013,Psaltis_2014,Zhao2024,Guver:2025wsj,Bootsma_2025}. One response is to adopt physics-informed hotspot parameterizations that reduce ad hoc geometric degrees of freedom; such approaches have recently been applied to PSR~J0030+0451 and PSR~J0437$-$4715 \citep{Huang:2025_hotspot}. It has also been shown by \citet{Lamb_2009,Lamb_2009b,Cole_2015} that for NICER-like data, several plausible sources of systematic error do not yield apparently acceptable fits with radius biases exceeding $1-\sigma$.

Second, PPM is computationally intensive: accurate forward models must couple relativistic ray tracing, realistic atmospheric beaming, rotation-induced oblateness, and instrumental background and calibration systematics. Rigorous Bayesian inference (e.g., nested sampling or ensemble Markov chain Monte Carlo, i.e., MCMC) typically requires $\sim 10^{6}$--$10^{8}$ likelihood evaluations in a high-dimensional parameter space \citep{2020MNRAS.493.3132S,2021JOSS....6.3001B} and incorporating radio constraints or joint NICER+XMM data further increases the cost \citep{Miller2021,Salmi_2024_j0740,Choudhury24}. Third, there is an accuracy-speed trade-off: coarse angular, phase, and energy grids can bias the likelihood and, thus, the posteriors for certain hotspot configurations; whereas the near-converged resolutions needed for robust inference (i.e., to render numerical errors subdominant to Poisson counting fluctuation) make forward modeling computationally demanding and can limit the scale of large-scale Bayesian analyses \citep{Choudhury_2024_waveform,Bogdanov_19,Bogdanov_21,Salmi_2024_Complex,Vinciguerra24}. Recent cross-code tests and case studies identify extreme geometries where insufficient resolution measurably alters the waveform and likelihood computation, motivating higher-resolution validation runs \citep{Choudhury_2024_waveform}. At present, the relatively large observational uncertainties from NICER mean that resolution choices do not yet have a visible effect on shifting the inferred mass-radius. Nonetheless, these differences are expected to become important for future missions such as {eXTP}, for which PPM is a key dense-matter science driver \citep{eXTP_Watts,eXTP_Li}. In practice, a single high-fidelity likelihood evaluation can take $\sim 10,\mathrm{s}$ or more (depending on resolution and model complexity), so many analyses adopt simplified hotspot parameterizations and/or multistage workflows (coarse-resolution exploration followed by high-resolution verification) to keep compute costs tractable. The principal bottleneck remains computational speed.

In this work, we address these limitations by introducing a GPU-accelerated PPM framework written in C++/CUDA. The framework supports energy- and phase-resolved data. NICER-specific instrument response and background models are implemented as modular components that can be swapped for other observatories. From the outset, the forward model accommodates {complex, physics-motivated} surface-temperature distributions beyond circular, isothermal caps, while preserving high-fidelity physics. Conceptually, our forward model adopts the standard physical ingredients established in prior CPU frameworks (e.g., \citealt{XPSI_joss}). Our contributions are: (i) a reengineered, GPU-optimized implementation of PPM for contemporary hardware; (ii) end-to-end reproducibility checks against existing benchmarks and codebases and a demonstration that the accuracy-speed trade-off can be broken by performing high-fidelity computation at unprecedented speed; and (iii) identification of a source of systematic error introduced by interpolating atmosphere lookup tables near grid boundaries.

Both the CPU and GPU implementations are released as open source to provide a reproducible foundation for future studies. In validation tests, the framework reproduces published waveform benchmarks to within $\lesssim 10^{-3}$ relative error while delivering $10^{3}$--$10^{4}\times$ speedups over representative CPU baselines. This reduces a single likelihood evaluation from seconds-minutes to milliseconds on a single GPU, making full Bayesian inference, which was previously practical only on large clusters, feasible on personal hardware.

The increased computation speed enables high angular, phase, and energy resolution without a prohibitive cost, mitigating the resolution-induced likelihood biases noted in prior work \citep{Choudhury_2024_waveform}. Consequently, a much broader family of fine-detail hotspot maps becomes computationally tractable, including cases previously impractical due to compute budgets. We anticipate that this framework will facilitate more realistic geometry priors, wider parameter exploration, and multi-instrument or multiwavelength joint analyses in forthcoming PPM studies built on this codebase.

The paper is organized as follows. Section~\ref{sec2:model_description} summarizes the theoretical framework, highlights subtleties that affect numerical accuracy, presents a transparent computational recipe to facilitate full reproducibility, and details our GPU implementation. Section~\ref{sec3:OS_tests} provides the first set of validation tests, demonstrating that our CPU and GPU implementations reproduce theoretical benchmarks under standard OS assumptions (see \citealt{Bogdanov_19}). Section~\ref{sec4:atmosphere_interpolation} investigates interpolation within atmosphere tables, detailing our interpolation strategy and introducing two new test cases that underscore the need for careful treatment. Section~\ref{sec5:production_level_result} reports production-grade computations that incorporate instrument response, interstellar absorption, and realistic atmospheric beaming (i.e., all the essential ingredients of PPM forward computation) and shows that the framework remains fast and robust even for hotspot configurations previously regarded as impractical, thereby overcoming the long-standing accuracy-speed trade-off. Section~\ref{sec6:discussion} concludes with a brief summary and discussion. Appendices~\ref{AppendixA:doppler}--\ref{AppendixC:flux_formula} supplement the main text with rederivations of some core formulas useful for PPM forward computation.

\section{Model framework and computational recipe}
\label{sec2:model_description}
In this section, we present the theoretical framework for X-ray PPM forward computation and provide a transparent, step-by-step recipe that covers the theoretical framework and implementation details to enable end-to-end reproducibility. The objective of PPM inference is to infer fundamental neutron-star properties from the energy- and phase-resolved X-ray waveforms. Given the statistical precision of modern data, we aim to limit forward-model inaccuracies to $\lesssim 1\%$ in phase-resolved flux where feasible.

The relevant physics combines general relativity (GR) and special relativity (SR). In GR, strong gravity near the neutron star (NS) alters photon trajectories and energies: gravitational light bending increases the visible surface fraction, revealing regions beyond the local limb; gravitational redshift reduces photon energies; and time-of-flight delays in curved spacetime introduce phase shifts. Rotation adds SR effects: Doppler boosting modulates the observed flux as a hotspot approaches or recedes from the observer, while rapid spin induces stellar oblateness. These effects are set by the global parameters $(M,R,\nu_*)$ and by the temperature map on the neutron-star surface. Foundational works include \citet{Pechenick_1987,Beloborodov_2002,Poutanen_2003,Miller1998,Poutanen_2006,Morsink_2007,Cadeau_2007,Psaltis_2014,Watts_2016,Nattila_2018,Poutanen_2020,zavlin_96,Ho_2001,Ho_2003,Suleimanov_2012,Salmi_2020,Salmi_2023,AlGendy_14,Viironen_04,Bogdanov_2008}.

The theoretical waveform is then propagated through the astrophysical and instrumental layers that connect the source to the data, namely, interstellar absorption and the instrument response. We address each component explicitly in dedicated subsections.

Our approach builds on modeling practices summarized by \citet{Bogdanov_19,Bogdanov_21,Salmi_2023} and the open-source \texttt{X-PSI} package \citep{XPSI_joss}, which we use as standard benchmarks. We extend these frameworks in a self-contained manner by providing fully relativistic derivations and implementations, applying controlled approximations only when they improve efficiency without biasing inference.

For orientation, Section~\ref{sec2:model_description} is organized as follows. Subsection~\ref{sec:geom} defines the physical setup and notation (parameters and geometry). Subsection~\ref{sec:os} specifies the spacetime metric and the oblate-surface model. Subsection~\ref{sec:light_bending} details photon propagation (light bending and time-of-flight delays). Subsection~\ref{sec:rotation} incorporates the rotational SR Doppler effect. Subsection~\ref{sec:beaming} models the surface emission and atmospheric treatment, integrating to obtain the observed flux. Subsection~\ref{sec:discretization} discusses the default surface temperature map deployed in this work and the surface discretization strategy. Subsection~\ref{sec:interstellar} discusses interstellar absorption and instrument response. Subsection~\ref{sect:computation_recipe} provides the detailed road map of our computation recipe, and the GPU implementation is discussed in Subsection~\ref{sec:GPU}. The objective is a production-grade, ready-to-use codebase for future PPM studies.

\subsection{Geometrical setup}
\label{sec:geom}
We modeled a rotating neutron star of mass $M$, equatorial radius $R_{\rm e}$, and spin frequency $\nu_*$ (angular frequency of $\Omega \equiv 2\pi \nu_*$). At the equator of a typical NS, the dimensionless surface speed is
\begin{equation}
\beta_{\rm eq} \equiv \frac{v_{\rm eq}}{c} \approx 0.1\text{--}0.3,
\end{equation}
so both GR and SR effects are necessary for sub-percent-level waveform accuracy.

Photons emitted from the oblate stellar surface experience light bending, gravitational redshift, time-of-flight delays, and Doppler boosting. To place these ingredients in a coherent framework, we adopted a star-centered coordinate system with the observer at infinity along a fixed line of sight. The stellar surface was discretized into small elements (i.e., surface patches) and we traced photons from each patch to the observer, integrating the contributions over the surface. Rather than restricting emission to one or a few circular, isothermal spots, our algorithm is designed from the outset to accept {full-surface temperature maps}, making it suitable for nontrivial configurations (e.g., accretion-heated distributions or physics-motivated hotspot maps). This relaxes symmetry assumptions that are typical for spot-based models; consequently, we emphasize computational optimizations to ensure the tractability of full-surface maps.

\begin{figure}
\centering
\includegraphics[width=0.6\linewidth]{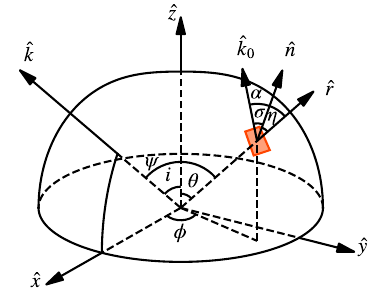}
\caption{Demonstration of the geometric setup used in our X-ray PPM forward computation, shown in the lab-frame coordinate system $\hat{x}-\hat{y}-\hat{z}$. A surface patch (orange square) on the stellar surface at colatitude $\theta$ and rotational phase $\phi$, viewed along the observer’s line-of-sight unit vector, $\hat{k}$. Definitions of all unit vectors are given in Table~\ref{tab:vectors} and angle conventions are summarized in Table~\ref{tab:angles}.}
\label{fig:geometry_setup_angles}
\end{figure}

Our geometric notation follows \citet{Loktev_2020} for consistency with prior literature. A schematic of the setup is shown in Figure~\ref{fig:geometry_setup_angles}. For clarity, the key unit vectors and angles used in this work are summarized in Tables~\ref{tab:vectors} and \ref{tab:angles}.

Physically, $i$ denotes the observer’s inclination; $(\theta,\phi)$ specify the location of an emitting surface element; $\alpha$ is the emission angle measured from the local radial direction in the near-star static frame; and $\psi$ is the angle between the surface radial vector and the observer’s line of sight. In GR, regions with $\psi>\pi/2$ remain visible due to gravitational light bending; the maximum visible $\psi$ is determined by the compactness (defined below).

\begin{table*}
\centering
\caption{Unit vectors in the geometric setup}
\label{tab:vectors}
\begin{tabular}{ll}
\hline
Symbol & Description \\
\hline
$\hat{{k}}$ & Direction from the star center to the distant observer. \\
$\hat{{r}}$ & Direction from the star center to the emitting surface patch. \\
$\hat{{n}}$ & Local surface normal at the emitting patch (coincides with $\hat{{r}}$ for spherical stars). \\
$\hat{{k}}_0$ & Initial photon direction at emission (before propagation). \\
$\hat{{z}}$ & Stellar rotation axis. \\
$\hat{{\beta}}$ & Local velocity direction of the emitting patch (tangent to the azimuthal circle). \\
\hline
\end{tabular}
\end{table*}

\begin{table*}
\centering
\caption{Angles in the geometric setup}
\label{tab:angles}
\begin{tabular}{ll}
\hline
Symbol & Description \\
\hline
$i$ & Inclination: angle between $\hat{{z}}$ and $\hat{{k}}$. \\
$\theta$ & Colatitude: angle between $\hat{{z}}$ and $\hat{{r}}$. \\
$\phi$ & Azimuth of $\hat{{r}}$ in the star frame. \\
$\psi$ & Angle between $\hat{{r}}$ and $\hat{{k}}$, with $\cos\psi = \hat{{r}} \cdot \hat{{k}} = \sin i \sin\theta \cos\phi + \cos i \cos\theta$. \\
$\eta$ & Angle between $\hat{{r}}$ and $\hat{{n}}$ (zero for spherical stars; nonzero for oblate stars). \\
$\alpha$ & Angle between $\hat{{r}}$ and $\hat{{k}}_0$ (emission angle relative to the radial direction). \\
$\sigma$ & Angle between $\hat{{n}}$ and $\hat{{k}}_0$ (emission angle relative to the local normal). \\
$\xi$ & Angle between $\hat{{\beta}}$ and $\hat{{k}}_0$; $\cos\xi \equiv \hat{{\beta}}\cdot \hat{{k}}_0 = -\,{\sin\alpha\,\sin i\,\sin\phi}/{\sin\psi}$. \\
\hline
\end{tabular}
\end{table*}

For a spherical star the surface normal coincides with the radial unit vector, $\hat{ n}=\hat{ r}$; hence,  $\eta=0$ and $\sigma=\alpha$. For an oblate star, rotational flattening tilts the surface normal away from $\hat{r}$; hence, $\eta\neq 0$. We quantify this with the dimensionless surface-slope parameter,
\begin{equation}
\label{eq:def_f}
f \equiv \frac{1}{R\,\sqrt{1-u}}\,\frac{dR}{d\theta}, \qquad u \equiv \frac{R_{\rm S}}{R} ,
\end{equation}
with $R_{\rm S}\equiv 2GM/c^2$ the Schwarzschild radius. We have
\begin{equation}
\cos\eta = \hat{{r}} \cdot \hat{{n}} = \frac{1}{\sqrt{1+f^2}}, 
\qquad 
\sin\eta = \frac{f}{\sqrt{1+f^2}} .
\end{equation}
The emission angle with respect to the local normal is
\begin{equation}
\label{eq:sigma_relation}
\begin{split}
\cos\sigma
=\, & \cos\alpha\,\cos\eta + \frac{\sin\alpha\,\sin\eta}{\sin\psi} \\
& \times \bigl(\cos i\,\sin\theta - \sin i\,\cos\theta\,\cos\phi\bigr).
\end{split}
\end{equation}

These definitions lay the foundation for computing the observed flux. This subsection provides a robust geometric framework that connects the static surface description with the fully relativistic treatment used in our forward model.

\subsection{Spacetime and oblate-Schwarzschild (OS) approximation}
\label{sec:os}
We go on to specify the exterior spacetime and the rotation-induced stellar shape. Throughout this work, we adopted the oblate-Schwarzschild (OS) approximation: we traced the photon geodesics in the Schwarzschild spacetime of mass, $M$, while representing the stellar surface as rotationally flattened using the empirical shape prescription of \citet{Morsink_2007}, as refined by \citet{AlGendy_14} and \citet{Silva:2020oww}. This extends the classic ``Schwarzschild+Doppler'' (S+D) treatment \citep{Poutanen_2006} by including surface oblateness, while neglecting frame dragging and the external quadrupole moment of the metric. The Schwarzschild exterior metric is expressed as
\begin{equation}
\label{eq:schwarzschild}
\begin{split}
ds^2 = -\left(1 - \frac{R_{\rm S}}{r}\right) & c^2\, dt^2 
+ \left(1 - \frac{R_{\rm S}}{r}\right)^{-1} dr^2 \\
+ \, & r^2 d\theta^2 + r^2 \sin^2\theta\, d\phi^2 .
\end{split}
\end{equation}
The $g_{tt}$ component determines gravitational redshift and time dilation, while spatial curvature bends null geodesics.
For a spherical star, the stellar radius is constant, $R=R_{\rm e}$, while the {comoving}-frame area element is
\begin{equation}
dS' = R_{\rm e}^{2}\, d\cos\theta\, d\phi .
\end{equation}
Rotation deforms the star because centrifugal support reduces the effective gravity near the equator. We modeled the radius profile as
\begin{equation}
\begin{split}
\label{eq:Rtheta}
R =& \,R_{\rm e}\,\bigl(1 + o_2 \cos^2\theta\bigr), \\
\qquad 
o_2 =&\, \bar{\Omega}^2 \bigl(-0.788 + 1.030\,x\bigr),
\end{split}
\end{equation}
where $o_2$ is the shape coefficient and the dimensionless frequency, $\bar{\Omega}$, and compactness parameter, $x$, are defined as
\begin{equation}
\label{eq:Omega_c_def}
\bar{\Omega} \equiv 2\pi \nu_{*}\sqrt{\frac{R_{\rm e}^3}{GM}}, 
\qquad 
x \equiv \frac{GM}{c^2 R_{\rm e}} .
\end{equation}
The quadratic term encodes polar flattening ($o_2<0$) and remains accurate for $\nu_{*}\lesssim 700\,\mathrm{Hz}$ and $R_{\rm S}/R_{\rm e}\lesssim 1$, with fractional errors $\lesssim 1\%$ when higher-order terms are neglected. For oblate star, the corresponding surface area element is expressed as
\begin{equation}
\label{eq:area_oblate}
dS' = R^2 \sqrt{1+f^2}\, d\cos\theta\, d\phi ,
\end{equation}
with $f$ defined in Equation~(\ref{eq:def_f}). This ensures that the element measures the proper area of the tilted surface. The effective surface gravity varies with latitude due to the combined effects of oblateness and rotation. Following \citet{AlGendy_14,Bogdanov_19}, we have
\begin{equation}
\begin{split}
g(\theta)/g_0 = 1 &+ \sin^2\theta \bigl[ \bar{\Omega}^2 (-0.791 + 0.776x) \\
&\quad + \bar{\Omega}^4 (-1.315x + 2.431x^2) + \bar{\Omega}^6 (-1.172x) \bigr] \\
&+ \cos^2\theta \bigl[ \bar{\Omega}^2 (1.138 - 1.431x) \\
&\quad + \bar{\Omega}^4 (0.653x - 2.864x^2) + \bar{\Omega}^6 (0.975x) \\
&\qquad - \bar{\Omega}^4 (13.47x - 27.13x^2) \bigr] \\
&+ |\cos\theta|\, \bar{\Omega}^4 (13.47x - 27.13x^2).
\end{split}
\end{equation}
Here, the reference gravity $g_0$ of the corresponding spherical star is 
\begin{equation}
g_0 = \frac{GM}{R_{\rm e}^2}\,\frac{1}{\sqrt{1-R_{\rm S}/R_{\rm e}}}.
\end{equation}
At the equator ($\theta=\pi/2$), gravity is reduced, potentially affecting atmospheric structure and beaming, whereas at the poles it is enhanced.

The OS approximation is well suited to the millisecond pulsars targeted by NICER (typical spins $\nu_{*}\lesssim 400~\mathrm{Hz}$) and delivers $\lesssim 1\%$ waveform accuracy in the tests we consider. Errors remain modest up to $\sim 600$--$700$~Hz, which motivates recent refinements of the OS approximation \citep{jakab2025gravitationalredshiftrapidlyrotating}. At higher spins ($\gtrsim 700$--$800$~Hz), frame dragging and the external metric’s quadrupole terms become non-negligible, and Hartle--Thorne or fully relativistic rotating-star spacetime might be required. These choices fix the geometric setting for the rest of the calculation: null geodesics in Schwarzschild provide the mapping from the local emission direction and emission time at the surface to the observed angle and phase, naturally incorporating gravitational light bending and time-of-flight delays.

\subsection{Photon propagation: light bending and time delays}
\label{sec:light_bending}

The dominant GR effects relevant to PPM are gravitational light bending, gravitational redshift, and time-of-flight delays in the curved spacetime around the neutron star. Gravitational light bending deflects photon trajectories away from straight lines, increasing the visible fraction of the surface; regions with $\psi>\pi/2$ that would be hidden in Newtonian geometry can remain visible. Time-of-flight delays introduce phase shifts between photons emitted at different surface locations and directions, modifying the waveform shape and its harmonic content. Accurate treatment of these effects is essential because modeling errors can bias the inferred compactness $GM/(Rc^{2})$ and hence the derived $M$ and $R$, rather than merely inflating statistical uncertainties. 

Gravitational light bending is quantified by the deflection angle $\psi$ for a photon emitted at radius, $R$, with the emission angle, $\alpha$, measured from the local radial direction. For outgoing rays ($0 \le \alpha < \pi/2$), it is given by the Schwarzschild null-geodesic integral,
\begin{equation}
\psi(u, \alpha) = \int_{R}^{\infty} \frac{dr}{r^2} \left[ \frac{1}{b^2} - \frac{1}{r^2} \left(1 - \frac{R_{\rm S}}{r} \right) \right]^{-1/2},
\label{eq:psi_integral}
\end{equation}
where the impact parameter is $b = {R \sin \alpha}/{\sqrt{1 - u}}$ and $u = R_{\rm S} / R$. This integral derives from the null geodesic equation in the Schwarzschild metric, reflecting how gravity pulls photons inward.

For inward emission ($\alpha > \pi/2$), the photon path passes through periastron before escaping via
\begin{equation}
\psi(u, \alpha) = 2\psi_{\text{max}} - \psi(u, \pi - \alpha),
\end{equation}
where $\psi_{\text{max}} \equiv \psi(u_p, \pi/2)$ and the periastron compactness is $u_p = u / \rho$, with
\begin{equation}
\rho =   \frac{-2 \sin \alpha}{\sqrt{3(1 - u)}} \cos\left( \frac{1}{3} \left[ \arccos\left( \frac{3u \sqrt{3(1 - u)}}{2 \sin \alpha} \right) + 2\pi \right] \right) .
\label{eq:rho}
\end{equation}

Using Equations~(\ref{eq:psi_integral})--(\ref{eq:rho}), we precomputed two-dimensional (2D) lookup tables that map $(u,\cos\psi)\mapsto\bigl(\cos\alpha, d\cos\alpha/d\cos\psi\bigr)$; these speed up ray tracing while retaining a high level of accuracy.

To calculate the time delay for a spherical star, we define zero delay for a photon emitted from the point directly beneath the observer (along $\hat{k}$). The gravitational and geometric delay for a photon emitted at the same radius $R$ but at angle $\alpha<\pi/2$ is
\begin{equation}
\label{eq:tp_direct}
\frac{c\,\Delta t_{p}}{R}(u,\alpha)
=
\frac{1}{R}\int_{R}^{\infty}\frac{dr}{1-R_{\rm S}/r}
\left\{
\left[
1-\frac{b^{2}}{r^{2}}\left(1-\frac{R_{\rm S}}{r}\right)
\right]^{-1/2} - 1
\right\}.
\end{equation}
For an oblate star, an additional delay arises from the difference between the actual emission radius, $R$, and the equatorial radius, $R_{\rm e}$,
\begin{equation}
\label{eq:delta_t_oblate}
c\,\delta t (R )
=
R_{\rm e} - R
+
R_{\rm S} \ln\left(\frac{R_{\rm e}-R_{\rm S}}{R-R_{\rm S}}\right).
\end{equation}
For direct emission ($\alpha<\pi/2$), the total delay is $\Delta t=\Delta t_{p}+\delta t$. For inward emission ($\alpha>\pi/2$), the photon first moves inward and then outward; the total delay is 
\begin{equation}
\label{eq:tot_time_delay}
\begin{split}
\frac{c\,\Delta t}{R}(u,\alpha)
=
\frac{c}{R}\Bigl[\,2\,\Delta t_{p}(u_{p},\pi/2)-\Delta t_{p}(u,\pi-\alpha)\\
+2\,\delta t(R_{p})-\delta t(R)\Bigr],
\end{split}
\end{equation}
where $R_{p}=\rho R$ is the periastron radius. Precomputing $(u,\cos\alpha)\mapsto (c\,\Delta t/R)$ yields fast, stable phase corrections for rotating stars. For a detailed explanation and discussion about the derivation above, we refer to \citet[Section~2.1 and Appendix~A]{Salmi2018}. Interpolating from these three tables during likelihood evaluations avoids repeated geodesic integrations and is therefore essential for Bayesian inference, which typically requires $\sim 10^{6}$--$10^{8}$ forward-model evaluations.

\subsection{Rotation: Relativistic Doppler effect}
\label{sec:rotation}
We incorporate rotation by superimposing SR surface effects on the full GR propagation described above: we Lorentz-boosted the emitted photons in {comoving} frame into the local {static} frame on the oblate surface and then trace them along Schwarzschild null geodesics \citep[e.g.,][]{AlGendy_14,Bogdanov_19}. For millisecond pulsars with spins $\nu_{*}\sim 300~\mathrm{Hz}$ (typical equatorial $\beta_{\rm eq}\equiv v/c \sim 0.1$), this ``OS\,+\,Doppler'' treatment provides an excellent approximation, whereas higher spins may necessitate Hartle--Thorne or fully relativistic rotating-star metrics. Special-relativistic Doppler boosting generates pulse asymmetry and higher harmonics, which are crucial for recovering the hotspot geometry on the neutron-star surface.

The surface velocity, accounting for the gravitational-redshift factor in the metric, is
\begin{equation}
\beta = \frac{2\pi\nu_* R \sin\theta}{c\sqrt{1-u}}, \quad \gamma = \frac{1}{\sqrt{1-\beta^2}},
\end{equation}
where the $\sqrt{1-u}$ factor accounts for the gravitational-redshift correction to the rotating frequency. 

The relativistic Doppler factor 
\begin{equation}
\delta = \frac{1}{\gamma (1 - \beta \cos\xi)},
\end{equation}
modifies the observed photon energy, $E_{\text{obs}}$, in static frame relative to the emitted energy in the comoving frame, $E_{\text{emit}}$, as $E_{\text{obs}} = \delta E_{\text{emit}}$. This factor depends on the velocity, $\beta$, of the emitting region and the angle, $\xi$, of the velocity vector with respect to the emitting direction. It tends to amplify energy and intensity for photons from approaching patches ($\cos\xi > 0$) and diminishes them for receding ones. The angles in the {comoving} frame (where the patch is at rest) are transformed via Doppler aberration,
\begin{equation}
\mu = \cos\sigma' = \delta \cos\sigma, \quad \cos\alpha' = \delta \cos\alpha.
\end{equation}
Quantities with a prime ($'$) are measured in the {comoving} frame. A detailed derivation of the above result is presented in Appendix~\ref{AppendixA:doppler}.

The rapid rotation of the neutron star means that the emission from a surface hotspot will be continuously Doppler-shifted throughout the star's rotational phase. As the hotspot moves across the stellar disk towards the observer, its emission is progressively blueshifted, reaching maximum blueshift as it approaches the line of sight. Conversely, as it recedes, the emission becomes increasingly redshifted. This modulation of photon energy directly impacts the observed flux and the shape of the pulse profile. Therefore, any realistic model of a pulse profile must incorporate the relativistic Doppler effect to correctly interpret the observed light curves and accurately constrain fundamental neutron-star parameters.
These SR effects couple directly to the intrinsic atmospheric beaming, as discussed in the next section.

\subsection{Surface emission and atmospheric beaming}
\label{sec:beaming}
When photons escape from surface hotspots, a simple starting point is to model the emergent spectrum as a blackbody in the {comoving} frame. For quantitative inference, however, physically motivated neutron-star atmosphere models are required. Thin (likely accreted) hydrogen or helium layers produce angle- and energy-dependent specific intensities (``beaming'') and spectra that deviate from a blackbody spectrum, typically appearing harder and exhibiting energy- and composition-dependent limb darkening or brightening. For rotation-powered millisecond pulsars with magnetic fields $B \sim 10^{8}$--$10^{9}\,\mathrm{G}$, weakly magnetized hydrogen atmospheres are commonly adopted; whereas strongly magnetized models are required at higher fields. We followed this atmosphere-based approach throughout, drawing on established calculations and recent updates \citep{zavlin_96,Ho_2001,Ho_2003,Suleimanov_2012,Salmi_2020,Salmi_2023,Choudhury_2024_waveform}.

According to \citep[Equation 31]{Bogdanov_19}, the number flux of photons can be calculated as
\begin{equation}
\begin{split}
df(E) = &\,\sqrt{1-u}\,\delta^3\,\mu\,\frac{d\cos\alpha}{d\cos\psi}\,\frac{I(E',\mu)}{E} \\
&\times \frac{\gamma\,R^2\,\sqrt{1+f^2}\,d\cos\theta\,d\phi}{D^2},
\end{split}
\end{equation}
where $df$ denotes the photon count rate per unit detector area and per unit observed photon energy; $I$ is the atmospheric specific intensity; $E$ and $E'$ denote the observed and emitted photon energies, respectively; and $D$ is the distance between the star and the observer. A detailed derivation of this expression is provided in Appendix~\ref{AppendixC:flux_formula}. Integrating over all visible surface elements, $d\cos\theta\,d\phi$, yields the full pulse profile. 

For realistic beaming, we followed the procedure of \citet{Bogdanov_21}. The characteristic field strength at which magnetic effects become non-negligible in the atmosphere is $B_0 \equiv e^3 m_e^2 c/\hbar^3 \simeq 2\times 10^9\,\mathrm{G}$. Magnetic effects are negligible in this context because surface fields are smaller than $B_0$. We also fixed the composition and ionization. After accretion ceases, gravitational settling stratifies the outer layers, leaving the lightest element at the photosphere; for recycled MSPs, this is typically hydrogen, with helium as a plausible variant. Over the NICER band and at $T_{\rm eff}\sim(0.5$--$2)\times 10^6\,\mathrm{K}$, differences in angular beaming between H and He are modest (a few percent below $1\,\mathrm{keV}$ and at most tens of percent near $2\,\mathrm{keV}$). However, at grazing angles $\bigl(\mu\lesssim0.5,\ \sigma^{\prime}\gtrsim60^\circ\bigr)$ these differences can become substantial \citep[see][]{Salmi_2023}. We therefore adopted a pure, nonmagnetic H atmosphere as the baseline and treat a pure He atmosphere as a systematic variant. At these temperatures the neutral fraction is small and for $\mu\gtrsim0.5$ fully ionized opacities reproduce the beaming of partially ionized solutions to within a few percent over $0.5-2 \,\mathrm{keV}$. Accordingly, we use fully ionized models for $I_\nu(\mu)$. We note that non-local thermodynamic equilibrium (NLTE) and Compton corrections are also negligible in this regime (sub-percent below $2\,\mathrm{keV}$). These choices keep the beaming physics self-consistent with the magnetic-field assumption above and isolate composition and ionization effects to small, quantifiable systematics in PPM inference. Throughout, we adopt the standard assumption of a {deep-heated} atmosphere in which external energy is deposited below the photosphere. Alternative heating geometries can alter the emergent spectrum and beaming \citep[e.g.,][]{Baub_ck_2019,Salmi_2020} and full consistency with the magnetosphere would require modeling the return-current particle velocities and their stopping column depths, which we defer to future work for improvement.

To compute realistic beaming $I_\nu(\mu)$, we did not solve the radiative transfer during PPM inference. Instead, we used precomputed, high-resolution lookup tables of $\log(I\,T_{\rm eff}^{-3})$ as a function of $\big(\log T_{\rm eff},\,\log g,\,\mu,\,\log(E/kT_{\rm eff})\big)$. The hydrogen grid spans $\log T_{\rm eff}\in[5.1,6.8]$ (helium: $[5.1,6.5]$), $\log g\in[13.7,14.7]$, $\mu\in[10^{-6},\,1]$ with extra nodes near $\mu\simeq 0,1$, and $\log(E/kT_{\rm eff})\in[-1.3,2.0]$. Production analyses use the fully ionized H grid. An identically sampled He grid is propagated as a composition systematic. For the full details of this treatment, we refer to \citet{Bogdanov_21} for further discussion.

With the atmosphere model established, the near-surface radiative physics is specified. Two additional components complete the forward model used for data comparison: (i) the surface temperature map and its numerical discretization on the oblate star; and (ii) the propagation to the detector, including interstellar absorption and the instrument response. The next two subsections develop these ingredients in detail. We now transition from the GR+SR forward model to its observational implementation, which converts model photon fluxes into measured counts.

\subsection{Surface temperature maps and surface discretizations}
\label{sec:discretization}
Unlike most published PPM studies, which, for computational reasons, typically discretize only parametrized hotspot regions, our framework natively discretizes the {entire} stellar surface temperature map. Prior work has largely employed one- or two-spot models with simple geometries (e.g., circular caps) and single or dual temperature components. While effective for many applications, such prescriptions cannot capture spatially complex temperature structures within a spot or extended patterns across the surface. Full-surface discretization, in contrast, accommodates nontrivial temperature distributions and global morphologies, which are required in several physical scenarios (e.g., accreting neutron stars with extended heating patterns and nonuniform temperature distribution; see \citealt{Das_2022,Das_2025}).

In our codebase, the primary input is a user-specified {full-surface} map $T_{\rm eff}(\theta,\phi)$ defined on an oblate surface mesh in colatitude, $\theta$, and azimuth, $\phi$. The mesh resolution is user-controlled and can be increased until interpolation errors are sub-percent in the phase--energy domain.

For benchmarking with results from the literature, we also provide parameterizations that reproduce standard geometric hotspot models used in previous PPM works. For example, a single circular cap is defined by its center, $(\theta_{0},\phi_{0})$, angular radius, $\zeta$, and spot temperature ,$T_{\rm spot}$; all mesh elements with angular separation $\le \zeta$ from $(\theta_{0},\phi_{0})$ are assigned $T_{\rm spot}$, while elements outside the cap are set to $0$, indicating that no photons are emitted. The same strategy extends to antipodal two-cap configurations and other predefined shapes.

Our default physics-motivated generator for whole-surface temperature maps follows \citet{Huang:2025_hotspot}, which derives return-current heating patterns from a force-free magnetosphere with an off-centered dipole. In that study, the implementation used the \texttt{X-PSI} \citep{XPSI_2021,XPSI_joss} full-surface module and achieved per-map generation times of order one second on a CPU, sufficient for static tests and small-scale inference but not optimized for large-scale PPM inference and for testing different sampler setups for robustness. In the present framework, we adopt this physics-motivated hotspot module as the default for hotspot configurations in future PPM inference. For the full algorithmic details and validation of the hotspot configuration, see \citet{Huang:2025_hotspot}.

\subsection{Interstellar absorption and instrument response}
\label{sec:interstellar}
Photons emerging from the stellar surface are attenuated by interstellar absorption, which we model as a multiplicative factor applied to the intrinsic spectrum. Accounting for this energy-dependent attenuation is essential for linking theoretical pulse profiles to the observed data and enabling reliable parameter inference. In this work, we model line-of-sight attenuation with a multiplicative transmission factor applied to the emitted photon spectrum,
\begin{equation}
f_{\rm ISM}(E;N_{\rm H}) = \exp\left[-\sigma_{\rm ISM}(E)N_{\rm H}\right] f_0(E),
\end{equation}
where $f_0(E)$ is the source-frame spectral flux, $N_{\rm H}$ is the hydrogen column, and $\sigma_{\rm ISM}(E)$ is the energy-dependent cross section from the \texttt{tbnew} prescription (modern gas-phase abundances and grain physics) \citep{wilms_00}. For efficiency, we use a precomputed attenuation vector on the NICER energy grid at a fiducial $N_{\rm H}^{\rm ref}$ and exploit the linear scaling in the exponent to obtain transmission for arbitrary $N_{\rm H}$. The MSPs targeted for PPM have very low columns ($\sim 10^{20}\,\mathrm{cm^{-2}}$), so extinction is small over $0.3$--$2$~keV and, being phase independent, has negligible impact on geometric inferences; we nevertheless carry $N_{\rm H}$ as a nuisance parameter and propagate it through the likelihood.

Another component that connects the theoretical pulse profile to the observed data is the instrument response. Here, our instrument response follows the same strategy implemented in previous NICER works as a standard pipeline. Comparison to NICER data is performed in detector space by convolving the attenuated model spectrum with the instrument response,
\begin{equation}
C(N) = \int_0^\infty R(N,E)\, f_{\rm ISM}(E;N_{\rm H}) \, dE,
\end{equation}
where $C(N)$ is the predicted count rate in channel $N$ (counts\,s$^{-1}$) and $R(N,E)$ is the full response (RMF$\times$ARF), namely, the probability of detecting a photon of energy, $E$, in channel, $N$, multiplied by the effective area. Energy-resolved pulse profiles are generated by evaluating $f_0(E)$ from the {atmospheric} beaming tables for each visible surface element and phase bin, applying the ISM transmission, and convolving with $R(N,E)$. We verified our forward model by reproducing standard XSPEC results for simple spectra with the same response files and by cross-checking independent implementations. All production fits use the identical RMF+ARF set and channelization as the data to ensure consistency.

\subsection{Computation recipe}
\label{sect:computation_recipe}
In this section, we outline the step-by-step computational procedure for generating a phase-resolved spectrum from a rotating neutron star. The process (i) discretizes the stellar surface, (ii) precomputes gravitational light bending, (iii) computes the observed flux from each surface element, and (iv) convolves the result with models for interstellar absorption and the instrument response.

\subsubsection{Surface discretization}
To accurately model the emission from the star's surface, we first divide it into a finite number of small patches or pixels. For this, we employ the \texttt{HEALPix} (Hierarchical Equal Area isoLatitude Pixelization) scheme \citep{HealPix_paper}. This method is particularly advantageous because it divides the sphere into patches of equal area, ensuring that each patch contributes proportionally to the total flux and simplifying the geometric calculations. The resolution of this grid is controlled by the parameter $N_{\rm side}$. Let $N_\theta$ be the number of rings and $i_\theta=1,\ldots,N_\theta$ the ring index. Each ring is further discretized into $N_\phi$ azimuthal bins with $i_\phi=1,\ldots,N_\phi$. The center coordinates $(\theta,\phi)$ of each surface patch are computed from $(i_\theta,N_\theta,i_\phi,N_\phi)$ according to the chosen discretization scheme.

For $1 \leq i_\theta < N_{\rm side}$ and $1 \leq i_\phi \leq 4i_\theta$, we have
\begin{equation}
\cos\theta(i_\theta,i_\phi,N_{\rm side})=1-\frac{i_\theta^2}{3N_{\rm side}^2},
\end{equation}
\begin{equation}
\phi(i_\theta,i_\phi,N_{\rm side})=\frac{\pi}{2i_\theta}\left(i_\phi-\frac12\right).
\end{equation}
For $N_{\rm side} \leq i_\theta \leq 2N_{\rm side}$ and $1 \leq i_\phi \leq 4N_{\rm side}$, we have
\begin{equation}
\cos\theta(i_\theta,i_\phi,N_{\rm side})=\frac43-\frac{2i_\theta}{3N_{\rm side}}, 
\end{equation}
\begin{equation}
\phi(i_\theta,i_\phi,N_{\rm side})=\frac{\pi}{2N_{\rm side}}
\left(i_\phi-\frac{(i_\theta-N_{\rm side}+1)~\rm{mod}~2}{2}\right).
\end{equation}
The southern hemisphere is a mirror image of the northern hemisphere, i.e., for $2N_{\rm side} < i_\theta \leq 4N_{\rm side}-1$ and $1 \leq i_\phi \leq 4\min (4N_{\rm side}-i_\theta, N_{\rm side})$, we have 
\begin{equation}
\cos\theta(i_\theta,i_\phi,N_{\rm side})=-\cos\theta(4N_{\rm side}-i_\theta,i_\phi,N_{\rm side}), 
\end{equation}
\begin{equation}
\phi(i_\theta,i_\phi,N_{\rm side})=\phi(4N_{\rm side}-i_\theta,i_\phi,N_{\rm side}).
\end{equation}
In the \texttt{HEALPix} discretization scheme, the surface of the sphere is partitioned into $12N_{\rm side}^2$ equal-area patches, so each patch covers a solid angle of $d\cos\theta\, d\phi=\pi/(3N_{\rm side}^2)$.

In modeling complex emission geometries, we identify the surface patches that belong to the emitting region and designate them as {active patches}, whose total number is $N_{\rm active}$. This discretization can introduce a small scaling bias at sharp boundaries because the pixelization may slightly over- or under-represent the true emitting area. A more careful edge treatment could mitigate this bias and potentially reduce computational load, but given that performance is satisfactory under the current setup, we defer such refinements to a future code release.

\subsubsection{Lensing table generation}
As discussed in Section~\ref{sec:light_bending}, computing gravitational light bending and the associated time-of-flight delays for every surface patch is computationally expensive. To accelerate the calculation, we precompute three lookup tables: (i) the cosine of the emission angle, $\cos\alpha$; (ii) the lensing factor, $d\cos\alpha/d\cos\psi$; and (iii) the time delay, $\Delta t$.

These tables are indexed by grids defined over the key physical parameters. We use a helper function to define linearly spaced arrays:
\begin{equation}
\bm{A} = \mathrm{Linspace}(x_{\min}, x_{\max}, x_{\rm num}),
\end{equation}
to denote an array $\bm{A}$ of size $x_{\rm num}$, whose elements are given by
\begin{equation}
\bm{A}[i] = x_{\min} + (x_{\max}-x_{\min}) \frac{i}{x_{\rm num}-1},
\end{equation}
where $i=0,\ldots,x_{\rm num}-1$.
The grids for the compactness ($u$), the cosine of the final bending angle ($\cos\psi$), and the cosine of the emitting angle ($\cos\alpha$) are defined as
\[
\label{eq:grids_def}
\bm{u} = \mathrm{Linspace}(u_{\min}, u_{\max}, N_u),
\]
\[
\bm{c}_\psi = \mathrm{Linspace}(c_{\psi,\min}, c_{\psi,\max}, N_{c_\psi}),
\]
\[
\bm{c}_\alpha = \mathrm{Linspace}(c_{\alpha,\min}, c_{\alpha,\max}, N_{c_\alpha}) .
\]
For all calculations, we adopted the following empirically chosen values, selected to ensure all lookups remain within bounds for typical neutron-star parameters:
\[
u_{\min}=0.3, \quad u_{\max}=0.5, \quad N_u=256,
\]
\[
c_{\psi, \min}=-0.9, \quad c_{\psi, \max}=1.0, \quad N_{c_\psi}=512,
\]
\[
c_{\alpha, \min}=-0.2, \quad c_{\alpha, \max}=1.0, \quad N_{c_\alpha}=512 .
\]

For $\alpha<\pi/2$,
\begin{equation}
\psi(u,\alpha)=\frac{2\sin\alpha}{\sqrt{1-u}}
\int_0^1\frac{x\,dx}{\sqrt{\cos^2\alpha+x^2q}},
\label{eq:salmi_lf}
\end{equation}
\begin{equation}
q = \Bigl[2-x^2-u(1-x^2)^2/(1-u)\Bigr]\sin^2\alpha ,
\end{equation}
and for $\alpha>\pi/2$,
\begin{equation}
\psi(u,\alpha)=2\psi(u/\rho,\pi/2)-\psi(u,\pi-\alpha) .
\end{equation}

We created two 2D arrays, $\bm{C}_\alpha$ and $\bm{LF}$, both of a size $(N_u, N_{c_\psi})$, to store precomputed lookup tables for the cosine of the local emission angle $\cos\alpha$ and the lensing factor $d\cos\alpha/d\cos\psi$. For each $\bm{u}[i_u]$ in the array $\bm{u}$, we set $\alpha_{\rm cur}=0$, calculated $\psi_{\rm cur}=\psi(\bm{u}[i_u],\alpha_{\rm cur})$, and appended $(\alpha_{\rm cur},\psi_{\rm cur})$ to a list $\{(\alpha,\psi),\ldots\}$. We then increased $\alpha_{\rm cur}$ by $d\alpha=\pi/2048$ (an empirically chosen constant; it can be decreased for higher accuracy) until $\psi_{\rm cur} > \arccos c_{\psi,\min}$. With a simple transformation, we converted the list $\{(\alpha,\psi),\ldots\}$ to $\{(\cos\psi,\cos\alpha),\ldots\}$; using numerical differentiation, we obtain the list $\{(\cos\psi,d\cos\alpha/d\cos\psi),\ldots\}$. Then, for each $\bm{c}_\psi[i_{c_\psi}]$ in the array $\bm{c}_\psi$, we interpolated $\bm{c}_\psi[i_{c_\psi}]$ within these two lists to obtain the corresponding $\bm{C}_\alpha[i_u,i_{c_\psi}]$ and $\bm{LF}[i_u,i_{c_\psi}]$. Next, we moved on to the calculation of the time delay. For $\alpha < \pi/2$,
\begin{equation}
\frac{c\Delta t_p}{R}(u,\alpha)=
\frac{2\sin^2\alpha}{1-u}
\int_0^1 \frac{x\,dx}{\sqrt{\cos^2\alpha+x^2q}\,\bigl(1+\sqrt{\cos^2\alpha+x^2q}\bigr)} .
\label{eq:salmi_td}
\end{equation}
For $\alpha > \pi/2$,
\begin{equation}
\frac{c\Delta t_p}{R}(u,\alpha)=2\,\frac{c\Delta t_p}{R}(u/\rho,\pi/2)-\frac{c\Delta t_p}{R}(u,\pi-\alpha) .
\end{equation}
We used a 2D array $\bm{TD}$ of size $(N_u,N_{c_\alpha})$ to store the lookup table of time delay as a function of $(u,c_\alpha)$.
For each grid point $(\bm{u}[i_u], \bm{c}_\alpha[i_{c_\alpha}])$, we directly calculated the time delay using the integral above and stored the result, expressed as
\begin{equation}
\bm{TD}[i_u,i_{c_\alpha}] = \frac{c\Delta t_p}{R}\bigl(\bm{u}[i_u],\arccos \bm{c}_\alpha[i_{c_\alpha}]\bigr) .
\end{equation}

The light-bending integral (Equation~\ref{eq:salmi_lf}) and the time-delay integral (Equation~\ref{eq:salmi_td}) are given in \citet[Appendix~A]{Salmi2018}.

\subsubsection{Number flux calculation}
This step computes the observed flux from each surface patch and assembles the pulse profile. We used a widely employed {``shift-and-add''} technique: rather than simulating every active patch across all rotation phases, we first computed a detailed light curve for a single {pseudo-patch} located at $\phi=0$ on a given colatitude ring $\theta$. This template light curve was then phase-shifted and added to the total to account for the actual longitudes of the active patches on that ring.

For each ring indexed by $i_\theta$ at a colatitude, $\theta$, we determined which patches are active and collect the azimuthal angles, $\phi$, of the active patches into an array, $\bm{\phi}_a$, of a length ,$N_{\phi_a}$. If $N_{\phi_a}=0$, the ring contains no active patches and is skipped. Otherwise, we prepared four arrays to store the pseudo-patch light-curve data at $\phi=0$. These were reused later when accumulating contributions from the active patches:
\begin{equation}
\bm{F}[i_{\rm fp}],\quad
\bm{z}[i_{\rm fp}],\quad
\bm{\varphi}_o[i_{\rm fp}],\quad
\bm{\mu}[i_{\rm fp}],
\end{equation}
where $i_{\rm fp}=0,...,N_{\rm fp}$ and $N_{\rm fp}$ is the number of fine rotation-phase points, this parameter controls precision. When the pseudo-patch rotates to azimuth $\phi(i_{\rm fp})=2\pi\,i_{\rm fp}/N_{\rm fp}$, we perform the following for each $i_{\rm fp} \in [0,N_{\rm fp}]$:
\begin{enumerate}
\item Initialize invisible-range indices $i_{\rm inv\_min}=N_{\rm fp}$ and $i_{\rm inv\_max}=0$.
\item Compute $\cos\psi = \sin i \sin\theta \cos\phi + \cos i \cos\theta$.
\item Use the lookup table to obtain $\cos\alpha$ and the lensing factor $d\cos\alpha/d\cos\psi$.
\item Evaluate $\cos\sigma$ from Equation~(\ref{eq:sigma_relation}):
\begin{equation}
\begin{split}
\cos\sigma =& \,\cos\alpha\,\cos\eta + \frac{\sin\alpha\,\sin\eta}{\sin\psi}\,\\
&\times \bigl(\cos i\,\sin\theta - \sin i\,\cos\theta\,\cos\phi\bigr).
\end{split}
\end{equation}
\vspace{-0.5cm}
\item If $\cos\sigma<0$, the point is invisible (and if $i_{\rm fp}=0$, the entire ring is invisible). Update
$i_{\rm inv\_min}=\min(i_{\rm inv\_min},i_{\rm fp})$ and $i_{\rm inv\_max}=\max(i_{\rm inv\_max},i_{\rm fp})$.
Otherwise, set
\begin{equation}
\begin{split}
\bm{F}[i_{\rm fp}]&=\sqrt{1-u}\,\delta^3\,\cos\sigma'\,\frac{d\cos\alpha}{d\cos\psi}\,
\frac{\gamma\,R^2 \sqrt{1+f^2}\, d\cos\theta\, d\phi}{D^2},\\[4pt]
\bm{z}[i_{\rm fp}]&=\frac{1}{\delta\sqrt{1-u}},\\
\bm{\varphi}_o[i_{\rm fp}]&=\frac{\phi}{2\pi} + \nu_* \Delta t,\\
\bm{\mu}[i_{\rm fp}]&=\delta\cos\sigma.
\end{split}
\end{equation}
\end{enumerate}

If a ring is partially visible, we must handle the phase segment where it is hidden. For these invisible points ($i_{\rm inv\_min} \leq i_{\rm fp} \leq i_{\rm inv\_max}$), we set the flux and apparent emission angle to zero. To ensure continuity and avoid numerical artifacts, the redshift and observed phase are smoothly filled using linear interpolation between the values at the points of disappearance ($i_{\rm inv\_min}-1$) and reappearance ($i_{\rm inv\_max}+1$), namely,
\begin{equation}
\bm{F}[i_{\rm fp}]=0,\qquad \bm{\mu}[i_{\rm fp}]=0,
\end{equation}
\begin{equation}
\begin{aligned}
\bm{z}[i_{\rm fp}]
&= L\!\Bigl(
i_{\rm fp};\, i_{\rm inv\_min}-1,\, i_{\rm inv\_max}+1,\\
&\hspace{2.2em}\bm{z}[i_{\rm inv\_min}-1],\, \bm{z}[i_{\rm inv\_max}+1]
\Bigr),
\end{aligned}
\end{equation}

\begin{equation}
\begin{aligned}
\bm{\varphi}_o[i_{\rm fp}]
&= L\!\Bigl(
i_{\rm fp};\, i_{\rm inv\_min}-1,\, i_{\rm inv\_max}+1,\\
&\hspace{2.2em}\bm{\varphi}_o[i_{\rm inv\_min}-1],\, \bm{\varphi}_o[i_{\rm inv\_max}+1]
\Bigr).
\end{aligned}
\end{equation}
where $L$ denotes linear interpolation:
\begin{equation}
L(x;x_0,x_1,y_0,y_1)=y_0+(y_1-y_0)\,\frac{x-x_0}{x_1-x_0}.
\end{equation}

After preparing these calculations, we used the ``shift-and-add'' technique to accumulate all active patches' contributions to the total waveform. We defined the phase points to be output as an array, $\bm{\varphi}_{\rm out}$, of a size, $N_{\varphi_{\rm out}}$, the observed photon energies as an array, $\bm{E}_{o}$, of a size $N_{E_{o}}$, and the total output photon number flux as a matrix, $\bm{F}_{t}$, of a specific  size $(N_{\varphi_{\rm out}},N_{E_{o}})$. For each $\phi = \bm{\phi}_a[i_{\phi_a}]$ with $i_{\phi_a}=0,..., N_{\phi_a}-1$, each $\varphi = \bm{\varphi}_{\rm out}[i_{\varphi_{\rm out}}]$ with $i_{\varphi_{\rm out}}=0,...,N_{\varphi_{\rm out}}-1$, and each $E = \bm{E}_{o}[i_{E_{o}}]$ with $i_{E_{o}}=0,...,N_{E_{o}}-1$. Next, we can compute
\begin{equation}
\varphi_p=\bigl((\varphi+\tfrac{\phi}{2\pi}+1-\bm{\varphi}_o[0])\bmod 1\bigr) + \bm{\varphi}_o[0],
\end{equation}
\begin{equation}
\begin{split}
F_p&=\mathrm{Interp}(\varphi_p;\bm{\varphi}_o,\bm{F}), \\
z_p&=\mathrm{Interp}(\varphi_p;\bm{\varphi}_o,\bm{z}), \\
\mu_p&=\mathrm{Interp}(\varphi_p;\bm{\varphi}_o,\bm{\mu}),
\end{split}
\end{equation}
where $\mathrm{Interp}(x_0;\bm{x}, \bm{y})$ denotes building an interpolation function from arrays $\bm{x}$ and $\bm{y}$ and evaluating it at $x_0$. The contribution to the total flux is found by scaling, $F_p$, with the specific intensity of the atmospheric emission, $I(z_pE, \mu_p)$, which depends on the redshifted energy and the apparent emission angle. This contribution is then added to the final light-curve matrix, $\bm{F}_t$,
\begin{equation}
\bm{F}_t[i_{\varphi_{\rm out}},i_{E_{o}}] \mathrel{+}= F_p \,\frac{I(z_pE,\mu_p)}{E}.
\label{eq:add_Ft}
\end{equation}

\subsubsection{Interstellar absorption and instrument response}
The photon number flux calculated above is a theoretical quantity. To compare it with real data, we must model its interaction with the interstellar medium (ISM) and the response of the detector. We can define logarithmically spaced arrays using the helper function
\begin{equation}
\bm{A} =\mathrm{Logspace}(x_{\min},x_{\max},x_{\rm num}),
\end{equation}
which represents an array $\bm{A}$ of length $x_{\rm num}$ whose elements are given by
\begin{equation}
\bm{A}[i] = \exp\!\left(\ln x_{\min} + \bigl(\ln x_{\max}-\ln x_{\min}\bigr)\frac{i}{x_{\rm num}-1}\right),
\end{equation}
where $i=0,\ldots,x_{\rm num}-1$.

Throughout all calculations, we use the grid
\begin{equation}
\bm{E}_o = \mathrm{Logspace}(0.1~\mathrm{keV},\, 5.215~\mathrm{keV},\, N_{E_o}=128).
\end{equation}

We modeled the absorption by the ISM using a precomputed table from the \texttt{tbnew} model. The table provides the attenuation fraction as a function of energy for a reference hydrogen column density ($N_{\rm H}^{\rm ref} = 4 \times 10^{18}\,\mathrm{cm}^{-2}$). We used $\bm{col1}$ to represent the energy column in the table and $\bm{col2}$ to represent the attenuation-fraction column. Our simulation uses $N_{\rm H}=2\times10^{19}\,\mathrm{cm}^{-2}$. Assuming the optical depth scales linearly with $N_{\rm H}$, the new attenuation is the original value raised to the power of the ratio $N_{\rm H}/N_{\rm H}^{\rm ref}$ (i.e., $=5$). We created an array $\bm{ATT}$ by interpolating the table to our energy grid and applying this scaling; $\bm{ATT}$ is of a size denoted as $N_{E_o}$ and elements expressed as
\begin{equation}
\bm{ATT}[i_{E_o}]=\mathrm{Interp} \bigl(\bm{E}_o[i_{E_o}];\bm{col1},\bm{col2}\bigr)^{5}.
\end{equation}

Next, we convolve the signal with the instrumental response of the detector (e.g., NICER). This requires two standard data files: an ARF file and an RMF file. The ARF file provides the effective area of the detector across different energies and contains three columns that we denote $\bm{E}_l$, $\bm{E}_h$, and $\bm{A}_e$ all with size $N_e$. The RMF file describes how photons of a given true energy are recorded in different detector output channels and contains a matrix, $\bm{R}$, of a specific size $(N_e,N_c)$. We multiplied the effective area from the ARF with the RMF to obtain a fine-resolution response matrix, $\bm{R}_f$, with a specific size $(N_e,N_c)$:
\begin{equation}
\bm{R}_f[i_e,i_c]=\bm{A}_e[i_e]\times\bm{R}[i_e,i_c].
\end{equation}

We defined the array, $\bm{e}_i$, with a length, $N_{E_o}$, whose elements are given by
\begin{equation}
\bm{e}_i[j]=
\begin{cases}
1, & \text{if } i=j,\\
0, & \text{otherwise.}
\end{cases}
\end{equation}
Define the downsampling matrix $\bm{DS}$ with size $({N_{E_o}, N_e})$ by
\begin{equation}
\bm{DS}[i_{E_o},i_e]=\int_{\bm{E}_l[i_e]}^{\bm{E}_h[i_e]} \mathrm{Interp}\bigl(x;\bm{E}_o,\bm{e}_{i_{E_o}}\bigr)\,dx.
\end{equation}
Combining ISM attenuation, $\bm{ATT}$, the downsampling matrix, $\bm{DS}$, and the fine instrument response, $\bm{R}_f$, yields the final, comprehensive response matrix, $\bm{RSP}$. This matrix directly converts the theoretical, energy-resolved photon flux into the expected counts per detector channel. The resulting $\bm{RSP}$ has a specific size  $(N_{E_o},N_c)$ and is given by
\begin{equation}
\bm{RSP}[i_{E_o},i_c]=\sum_{i_e=0}^{N_e-1} 
\bm{ATT}[i_{E_o}] \times \bm{DS}[i_{E_o},i_e] \times \bm{R}_f[i_e,i_c].
\end{equation}

\subsection{GPU implementation}
\label{sec:GPU}

The NVIDIA GPU architecture is designed for massively parallel computation, executing thousands of threads simultaneously. The core processing unit is the {Streaming Multiprocessor} (SM), each of which contains hundreds of simpler arithmetic units known as CUDA cores. When a computational task is launched, it is organized into a grid of thread blocks; one or more blocks can be assigned to an SM for execution. The efficiency of this architecture depends on its memory hierarchy. All threads can access a large, high-capacity {global memory}, which is analogous to CPU random-access memory (RAM) but has high latency, making frequent access a primary performance bottleneck. To mitigate this, each SM is equipped with a small, extremely fast on-chip {shared memory}. This memory is private to the threads within a single block and acts as a programmable cache. In a typical GPU algorithm for a problem such as the one described below, a block of threads first cooperatively loads the necessary subset of data from global memory into this low-latency shared memory. All subsequent computations can then be performed by accessing the shared memory, reducing memory overhead and unlocking the full parallel processing power of the CUDA cores before writing the final results back to global memory.

In our waveform generation process (as described in Section~\ref{sect:computation_recipe}), we treated each ring independently. In the GPU implementation, we assigned each ring’s computation to a dedicated SM. For each SM, the four helper arrays $\bm{F}, \bm{z}, \bm{\varphi}_o,$ and $\bm{\mu}$ were staged in shared memory, whereas the number-flux matrix, $\bm{F}_t$, the lensing lookup table, the (optional) numerical atmosphere lookup table, and the $\bm{RSP}$ matrix reside in global memory. When a kernel is launched, the host (CPU) provides all configuration data describing the task, including the star’s mass, radius, and spin frequency ($M,R,\nu_*$); the observer’s inclination ($i$); the hotspot’s shape, position, and size ($\{\theta_i,\phi_i,\zeta_i,\ldots\}$); and the ring indices it operates on ($i_\theta,N_\theta$). For each active surface patch, output phase, and output energy, the kernel computes the corresponding number flux and immediately writes the result to the matrix $\bm{F}_t$ stored in global memory using Equation~\ref{eq:add_Ft}. After all tasks on the SMs were complete, we multiplied the matrix, $\bm{F}_t$, by the matrix, $\bm{RSP}$, which combines ISM absorption and the detector’s instrument response. We then copy the final result from GPU global memory to CPU host memory for analysis and plotting.

\section{OS approximation benchmarks}
\label{sec3:OS_tests}

In this section, we describe how we tested the OS approximation and reproduced the theoretical benchmarks established in \citet{Bogdanov_19} that were generated by the Illinois--Maryland (\texttt{IM}) group code. At this stage, no atmosphere model or instrument-specific modeling was included; all tests assumed blackbody emission. We therefore evaluated the code accuracy using blackbody spectra and prescribed beaming laws.

Following \citet{Bogdanov_19}, we verified both S+D and OS. The S+D scheme treats special-relativistic effects exactly at the surface, assuming a spherical star and using the exterior Schwarzschild metric; it is very fast and accurate for slowly rotating stars \citep[see also][]{Poutanen_2006}. The OS scheme incorporates the star’s spin-induced oblateness while retaining an exterior Schwarzschild spacetime; it is now the standard in X-ray PPM. Direct comparisons against ray tracing in numerically computed rotating-star metrics show that OS waveforms agree with the exact results to within $\lesssim 0.1\%$ for spin frequencies $\nu_{*}\lesssim 300~\mathrm{Hz}$, whereas S+D can incur percent-level errors at higher spins. Hence, we adopted OS as the default for our production runs \citep{Bogdanov_19,Morsink_2007} and implemented the OS approximation in our GPU codebase.

We followed the hotspot geometries and spacetime choices in \citet{Bogdanov_19}. Specifically, we tested both point-like ($\zeta_{\rm spot}=0.01~\mathrm{rad}$) and extended ($\zeta_{\rm spot}=1~\mathrm{rad}$) circular hotspots. Beaming patterns included (i) isotropic emission and (ii) simple parametric laws $\propto \cos^{2}\sigma$ and $\propto \sin^{2}\sigma$ used for verification in \citet[][Tables~3--4]{Bogdanov_19}, where $\sigma$ is the angle between the surface normal and the emission direction. In this section, we assume a canonical neutron-star mass $M=1.4\,M_\odot$, equatorial radius $R_{\rm e}=12~\mathrm{km}$, distance $D=200~\mathrm{pc}$, and a surface temperature $kT=0.35~\mathrm{keV}$. We considered monochromatic pulse profiles evaluated at $E_{\rm obs}=1~\mathrm{keV}$ with a photon number flux of $1~\mathrm{cm^{-2}\,s^{-1}\,keV^{-1}}$, matching the reference tests in \citet{Bogdanov_19}.

We implemented both CPU and GPU versions of our codebase. For validation, we compare our waveforms phase-by-phase against the \texttt{IM} reference waveforms used in \citet{Bogdanov_19}. Consistent with the NICER validation protocol, our target is a fractional accuracy better than $0.1\%$ across phases under likelihood-relevant weighting, where phase bins contribute in proportion to their expected counts. In this sense, we achieve the target accuracy, with larger relative deviations confined to ingress/egress when the hotspot lies behind the stellar limb and the flux is orders of magnitude below the pulse maximum. Because Poisson likelihood contributions scale with the expected counts, these low-flux bins carry little statistical weight in typical datasets. Moreover, the \texttt{IM} treatment about the stellar limb, as implemented in \citet{Bogdanov_19}, is a useful reference but not ground truth. Accordingly, we defer any certification of limb (small-$\mu$) influence on the waveform computation to a dedicated stress test in \ref{sec4:atmosphere_interpolation}, Since these systematic errors become significantly more consequential under realistic atmosphere model. These localized discrepancies have a negligible impact on inference because likelihood evaluations are generally insensitive to tiny, low-count portions of the profile.

\begin{figure*}
    \centering
    \includegraphics[width=1.0\linewidth]{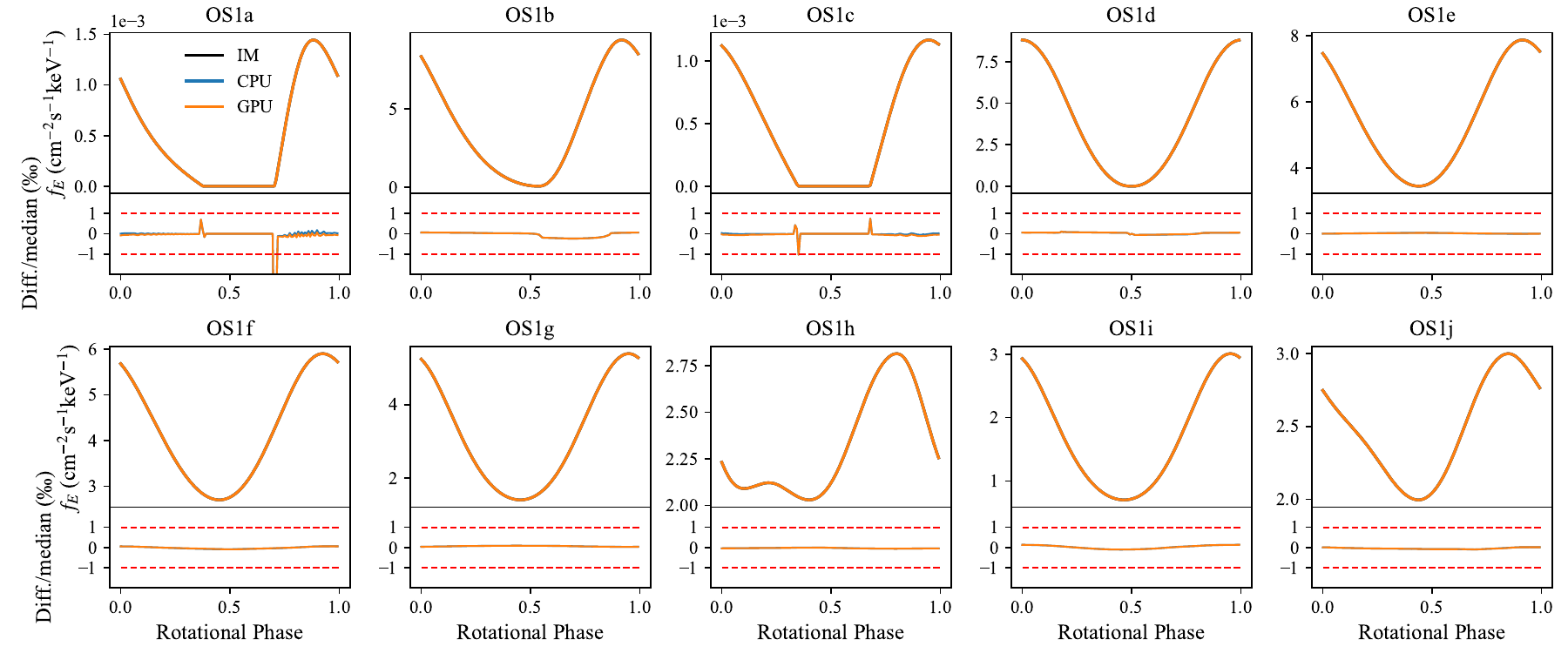}
    \caption{Monochromatic ($E_{\rm obs}=1~\mathrm{keV}$) waveforms under the oblate-Schwarzschild (OS) approximation for the OS1 test suite (subset shown), computed with our CPU (blue) and GPU (orange) implementations. The theoretical benchmarks from \citet{Bogdanov_19} are shown in black. Residuals are taken with respect to the benchmark; red dashed lines indicate $\pm 0.1\%$. Test-case definitions follow \citet{Bogdanov_19}.}
    \label{fig:OS_test}
\end{figure*}

Across all OS test cases (OS1a--OS1l; see Figure~\ref{fig:OS_test} for a subset), our CPU and GPU implementations reproduce the \texttt{IM} reference waveforms within the $0.1\%$ target at essentially all phases outside eclipse ingress/egress. The few outliers occur precisely where the spot rotates out of and back into view; as noted above, these phases have minimal photon counts and do not affect practical parameter estimation. This level of agreement meets the precision standard required for theoretical PPM forward computation and provides a reliable foundation for extensions such as energy-resolved profiles with {atmospheric} beaming and instrument responses. In the next section, we incorporate a realistic atmosphere model and identify an important interpolation artifact in the widely adopted atmospheric lookup table. We then introduce two diagnostic test cases to quantify its impact and present a more detailed implementation that mitigates the artifact.

\begin{figure*}
  \centering
  \begin{subfigure}{0.2\linewidth}
    \centering
    \includegraphics[width=\linewidth]{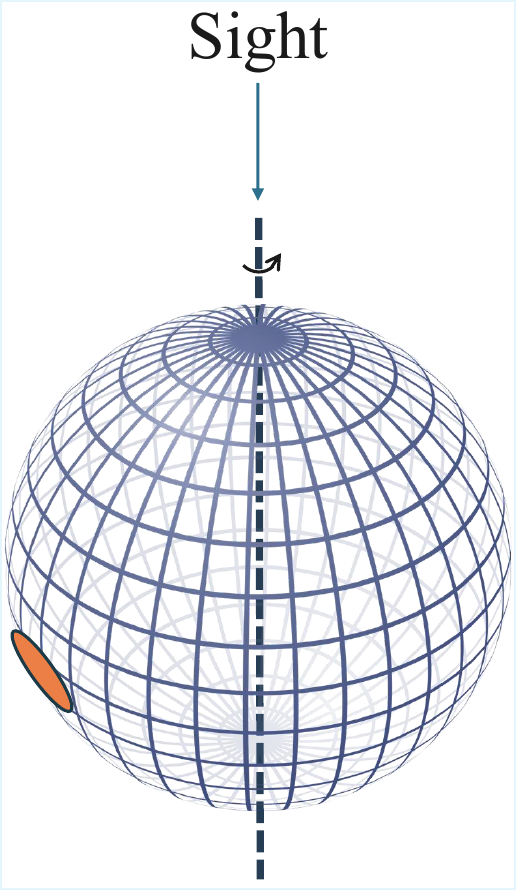}
    \caption*{Test 1}
  \end{subfigure} \hspace{0.1\textwidth}
  \begin{subfigure}{0.4\linewidth}
    \centering
    \includegraphics[width=\linewidth]{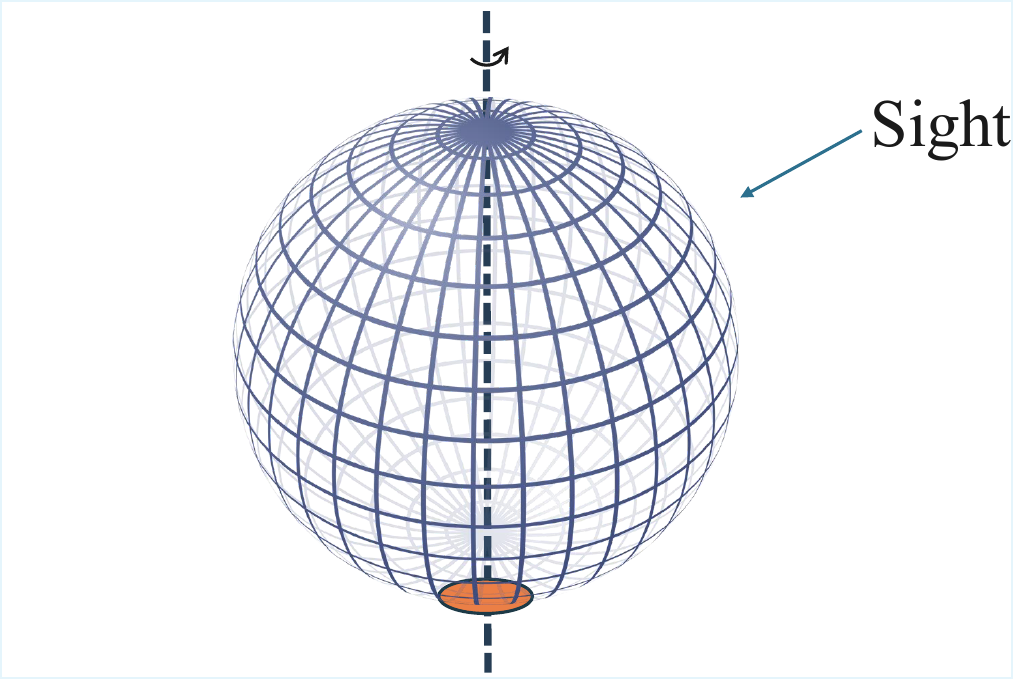}
    \caption*{Test 2}
  \end{subfigure}
\caption{Demonstration of two geometric test cases for atmosphere interpolation. Test 1: Point-like hotspot on the southern hemisphere; the line of sight is aligned with the spin axis and lies above the rotational north pole. Test 2: Point-like hotspot at the south pole; the observer is located in the northern hemisphere.}
\label{fig:demo_geometry}
\end{figure*}

\begin{figure*}
    \centering
    \includegraphics[width=1.0\linewidth]{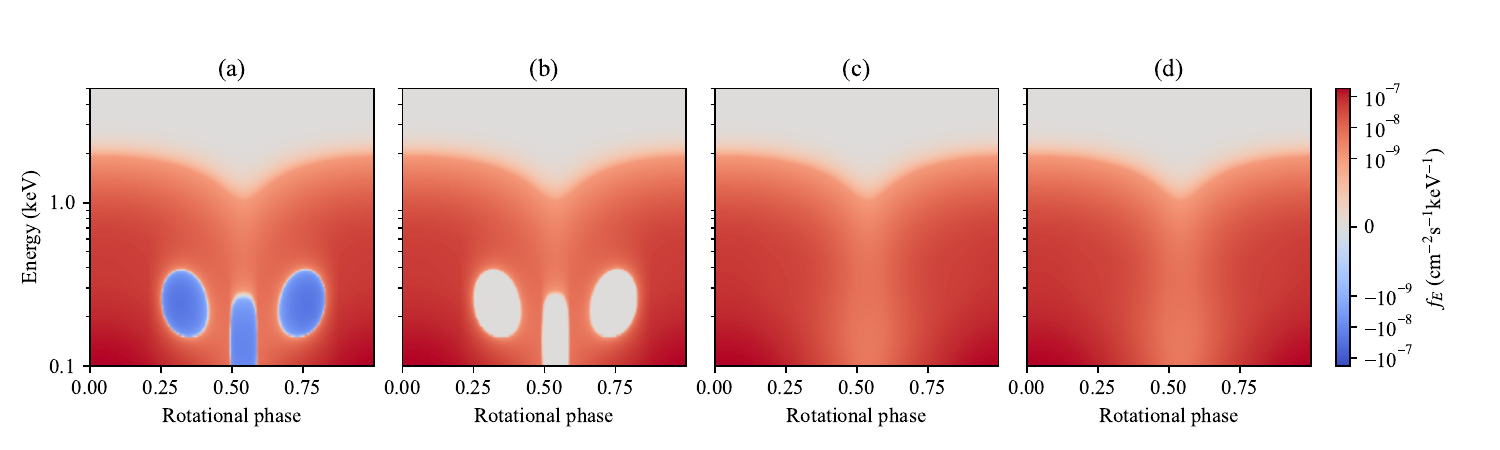}
    \caption{Simulated phase--energy--resolved X-ray pulse profiles for interpolation test 1, computed using four interpolation schemes applied to the NSX tables: {(a)} cubic interpolation in all dimensions; {(b)} as in (a), with negative intensities clipped to zero; {(c)} cubic interpolation in all dimensions with linear interpolation enforced at table boundaries; {(d)} the baseline configuration used in this work, cubic along two axes and linear along the remaining two, with linear interpolation enforced at table boundaries.}
    \label{fig:Test_interpolation1}
\end{figure*}

\begin{figure*}
    \centering
    \includegraphics[width=1.0\linewidth]{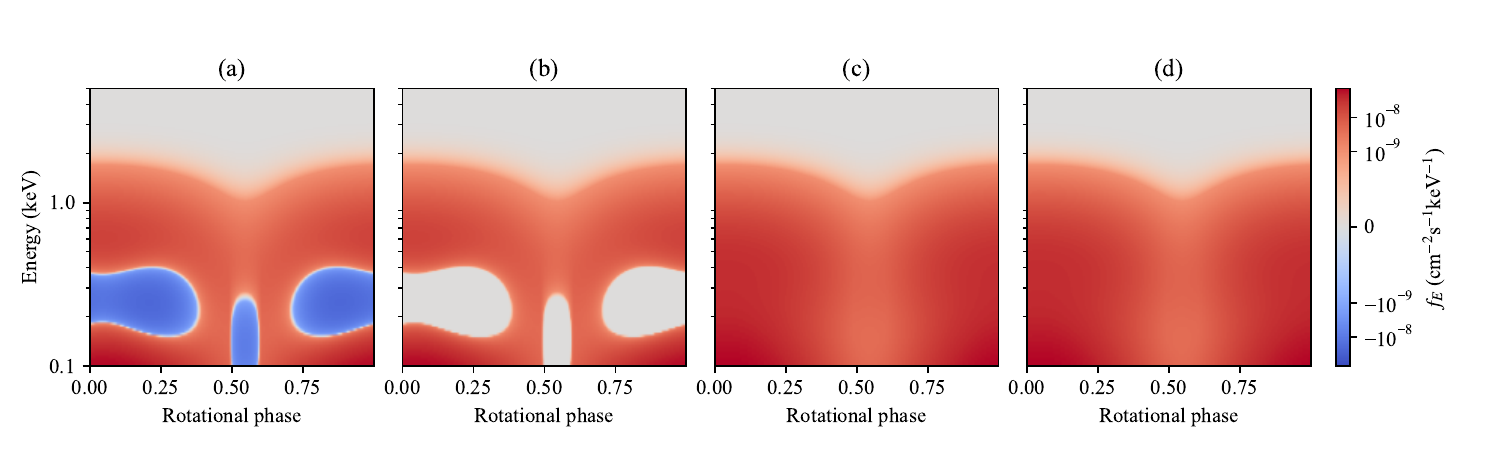}
    \caption{Simulated phase--energy--resolved X-ray pulse profiles for interpolation test 2, computed using four interpolation schemes applied to the NSX tables: {(a)} cubic interpolation in all dimensions; {(b)} as in (a), with negative intensities clipped to zero; {(c)} cubic interpolation in all dimensions with linear interpolation enforced at table boundaries; {(d)} the baseline configuration used in this work, cubic along two axes and linear along the remaining two, with linear interpolation enforced at table boundaries.}
    \label{fig:Test_interpolation2}
\end{figure*}
\section{Test cases on atmosphere interpolation}
\label{sec4:atmosphere_interpolation}
In the previous section, we explain how we verified that our implementation reproduces published theoretical benchmarks under the OS approximation and that the resulting light curves agree with canonical test problems. Before turning to production-grade studies, we highlight another common source of systematic error that is missing from much of the existing literature: interpolation within precomputed neutron-star atmosphere tables.

To avoid repeatedly solving the radiative-transfer problem on the fly, modern forward models use predefined lookup tables of the emergent specific intensity. This step is essential for realistic PPM: replacing a blackbody with a physically motivated atmosphere is required for meaningful comparison with NICER observations. In practice, many groups use nonmagnetized atmosphere grids for hydrogen or helium compositions, often referred to as \texttt{NSX-H} and \texttt{NSX-He}, which tabulate $I_\nu(\mu; T_{\rm eff}, g, E)$ as a function of photon energy $E$, emission-angle cosine $\mu=\cos\sigma'$, effective temperature $T_{\rm eff}$, and surface gravity $g$ \citep[e.g.,][]{Ho_2001,Ho_2003}. In what follows, we examine how interpolation choices on these grids can affect the modeled waveforms, and we propose additional test cases designed to demonstrate the numerical behavior of interpolation on the NSX atmosphere lookup tables. In our experiments, evaluating the NSX atmosphere tables and performing the associated interpolation constitute the dominant computational cost in pulse profile calculations. More importantly, different interpolation schemes can introduce distinct numerical artifacts (e.g., overshooting near table boundaries), which may bias predicted fluxes and spectra quite significantly under our proposed test cases.

Several widely used codes adopt different interpolation strategies for these tables. As reported in \citet{Bogdanov_19,Bogdanov_21,Choudhury_2024_waveform}, the \texttt{IM} code employs quadratic and quartic polynomial interpolation, \texttt{X-PSI} primarily uses cubic Lagrange interpolation, and the \texttt{Alberta} implementation relies on linear schemes (see \citealt{Bogdanov_19,Bogdanov_21,Choudhury_2024_waveform} for details). Because \texttt{IM} is not open source, its precise interpolation details are not publicly documented. In contrast, the publicly available \texttt{X-PSI} code uses cubic Lagrange interpolation for atmosphere lookups \citep{XPSI_joss,Bogdanov_21}. Given the structure of the tabulated NSX H/He atmosphere grids \citep{Ho_2001,Ho_2003}, these choices can introduce method-dependent artifacts, underscoring the need for careful treatment.

A commonly underappreciated numerical issue in atmosphere-table usage is {polynomial overshoot} near grid boundaries. Because the widely used nonmagnetized NSX H/He grids have finite spacing and regions of rapid curvature, selecting an interpolation scheme that avoids spurious oscillations while preserving physical constraints (e.g., non-negativity of the specific intensity) is nontrivial. In principle, as the grid becomes sufficiently dense, low-order (e.g., linear) interpolation is typically the most robust and economical. A comprehensive, controlled comparison of interpolation orders and monotone schemes, ideally alongside finer atmosphere grids or direct radiative-transfer recalculations, is deferred to future work.

In our experiments with \texttt{NSX-H} grids, cubic Lagrange interpolation can exhibit boundary instabilities (overshoot/undershoot) that depend on local spacing and curvature, particularly as the emission-angle cosine approaches an endpoint ($\mu \to 0$; grazing emission). To enforce non-negativity, the open-source \texttt{X-PSI} implementation clips any negative interpolants to zero.\footnote{\textsc{X-PSI} version 3.1.} While this safeguard prevents unphysical negative intensities, it still can't compute the predicted flux downward properly in phases where many rays sample boundary regions. 

Motivated by these considerations, we adopt a hybrid interpolation scheme in this work. For the temperature and gravity axes of the \texttt{NSX-H} table, $\log T_{\rm eff}$ and $\log g$ (the \texttt{NSX-H} tables tabulate these quantities in logarithmic units), we employ linear interpolation. The grids are sufficiently dense, making linear interpolation both robust and accurate in practice. For the emission-energy axis, we use cubic Lagrange interpolation in $\log (E/kT_{\rm eff})$; for the emission-angle cosine $\mu$, we use cubic Lagrange interpolation in the interior but switch to linear interpolation within boundary zones at $\mu \to 0$. Because a cubic Lagrange interpolant uses four points, we designate the outermost four grid nodes on the small-$\mu$ end as ``protected'' regions in which linear interpolation is enforced.

\begin{table*}
\centering
\caption{Parameter values for atmosphere-interpolation tests}
\label{tab:interp}
\begin{tabular}{lcc}
\toprule
{Case name} & {Test 1} & {Test 2} \\
\hline
Number of hotspots & 1 & 1 \\
Angular radius of hotspot (rad) & 0.01 & 0.01 \\
Colatitude of spot center $\theta_{\mathrm{c}}$ (rad) & 1.983 & $\pi-0.01$ \\
Observer colatitude $i$ (rad) & 0.01 & 0.964 \\
NS mass ($M_\odot$) & 1.4 & 1.4 \\
NS equatorial radius (km) & 12 & 12 \\
$\nu$ at infinity (Hz) & 600 & 600 \\
Spectrum of emission & \texttt{nsx\_H\_v200804} & \texttt{nsx\_H\_v200804} \\
Temperature of emission (K) & $10^6$ & $10^6$ \\
Distance (pc) & 150 & 150 \\
\hline\hline
\end{tabular}
\end{table*}

In a preliminary numerical test, we directly interpolated the \texttt{NSX-H} atmosphere tables and examined the resulting specific intensities. We found that the boundary behavior as $\mu\to0$ (i.e., near the stellar limb) is particularly susceptible to interpolation overshoot, occasionally producing unphysical negative intensities. This sensitivity arises from the steep angular dependence of the tabulated intensities close to the limb and the discrete structure of the \texttt{NSX-H} grids. To illustrate how such overshoot can affect otherwise typical observing configurations, we present two deliberately chosen test cases.

Both tests use a single circular hotspot with uniform temperature $T=10^6\,\mathrm{K}$, representing the simplest hotspot configuration. The hotspot has an angular radius of $0.01~\mathrm{rad}$ (i.e., a compact, highly localized patch), ensuring that viewing-geometry effects dominate, minimizing complex-shape-induced systematics and avoiding dilution from extended or complex emission regions. In both cases, the neutron-star mass and radius are fixed at $M=1.4\,M_{\odot}$ and $R=12\,\mathrm{km}$. To probe rapid rotation while remaining within the OS regime, we set the spin frequency to $600\,\mathrm{Hz}$. The same version of the \texttt{NSX-H} atmosphere tables is used throughout, and the source distance is fixed at $150\,\mathrm{pc}$. The colatitude of the hotspot center is $\theta_{\mathrm{c}}=1.983\,\mathrm{rad}$ for test~1 and $\theta_{\mathrm{c}}=\pi-0.01\,\mathrm{rad}$ for test~2; the observer colatitude is $i=0.01\,\mathrm{rad}$ for test~1 and $i=0.964\,\mathrm{rad}$ for test~2. A complete specification of both configurations is summarized in Table~\ref{tab:interp}, and their geometries are illustrated in Figure~\ref{fig:demo_geometry}.

The two tests differ only in viewing geometry and hotspot location. They are constructed to create more phases when the spot approaches or recedes beyond the visible limb, driving rays with $\mu\to 0$, where unconstrained cubic interpolation tends to overshoot. Although these examples intentionally maximize the visibility of the effect, the underlying issue is generic: for many reasonable geometries, a rotating hotspot will periodically sample small-$\mu$ rays (as it grazes the limb or re-enters the line of sight), making careful treatment near $\mu\to 0$ essential. 

Importantly, these setups are not exotic. Test~1 is consistent with a southern magnetic pole in a canonical inclined dipole or an off-centered dipole configuration. Test~2 is likewise plausible for an off-centered dipole whose south pole lies near the rotation pole. Variants such as a narrow ring hotspot at the same colatitude (e.g., arising from higher multipolar fields) can be analyzed analogously and would exhibit similar sensitivity to the limb region.

In each test case, we evaluated four strategies for interpolating the NSX table. Strategy {(a)} applies cubic interpolation uniformly across all four dimensions. Strategy {(b)} also uses cubic Lagrange interpolation but enforces a non-negativity safeguard by clipping negative values to zero. Strategy {(c)} employs cubic interpolation uniformly in all dimensions with linear interpolation enforced at the $\mu\to0$ boundary. Strategy {(d)}, our baseline in this work, uses linear interpolation in two dimensions and cubic interpolation in the remaining two, while reverting the nominally cubic dimensions to linear near the $\mu\to0$ boundary to suppress overshoot. All results are produced with the same CPU codebase and a high-resolution configuration.

Figure~\ref{fig:Test_interpolation1} presents phase--energy--resolved X-ray pulse profiles for test~1. For a small viewing angle ($i=0.01$), the profile exhibits mild asymmetry over spin phase. Under uniform cubic interpolation (panel~{a}), we observe pronounced negative intensities (large undershoots) centered near phases $\sim 0.25$, $0.50$, and $0.75$, predominantly in the soft X-ray band. These unphysical negatives would severely corrupt the inferred pulse profile and any derived likelihood. For the X-PSI code (which we have not run ourselves), a preliminary replication (T. Salmi, priv. comm.) reports a smaller negative flux (occuring at a different location) than we find here. We have not independently verified the phase–energy locations of these occurrences and therefore do not draw conclusions about where in the profile they arise.

Introducing non-negativity clipping (panel~{b}) removes the undershoots but creates extended zero-intensity “holes” that still end up biasing the likelihood computation by artificially suppressing flux. In contrast, uniform cubic interpolation with linear treatment at the edges (panel~{c}) yields a qualitatively reasonable profile: by construction, linear schemes at the boundaries strongly limit overshoot and thus avoid negative intensities. Our baseline mixed scheme (panel~{d}) likewise produces a physically consistent profile; the boundary reversion to linear in the nominally cubic dimensions effectively eliminates the negative regions, directly implicating cubic overshoot at the table edges of the $\mu$ axis as the root cause of these artifacts.

Test~2 (Figure~\ref{fig:demo_geometry}) is intentionally close in observables to test~1, but it is physically distinct: the hotspot is fixed at the rotational south pole while the line of sight changes which longitudes are viewed as the star rotates. This geometry is well motivated by centered or off-centered dipolar fields heating neutron-star hotspots. The same qualitative failure modes reappear under uniform cubic interpolation: three phase-localized negative regions in the soft band, which become zeroed “holes” after clipping, leading to substantial underestimation of photon counts. As before, both the uniform linear scheme and our mixed baseline recover stable, physically plausible profiles for this configuration.

Crucially, these cases are not isolated to a single demonstrated geometry. Small perturbations of the hotspot location, changes in mass--radius, or alternative hotspot shapes readily reproduce these negative regions when unconstrained cubic interpolation is used near table boundaries. Because Bayesian inference explores the full parameter space of a given model, it is effectively inevitable that such configurations will be encountered during inference. Unless interpolation is handled with care, mass--radius constraints risk hidden systematics and hence reduced credibility. We therefore recommend that existing pipelines incorporate these two test cases into their validation suites and adopt atmosphere-table interpolation strategies that control boundary overshoot (e.g., uniform linear or mixed schemes with boundary linearization, as implemented here). 

After this detailed discussion of the implementation of the atmosphere model, we  proceeded to production-grade tests. Specifically, we reproduced end-to-end computation cases that incorporate all necessary ingredients, such as our baseline atmosphere interpolation scheme, interstellar absorption, and convolution with the instrument response. These tests demonstrate that our treatment of the forward computation of the pulse profile is implemented correctly. With these components in place, the codebase is prepared for inference while achieving between three and four orders of magnitude speedup in pulse profile synthesis relative to existing CPU- and Python-based codebases.

\section{Production-level computation result}
\label{sec5:production_level_result}

In this section, we proceed by starting with theoretical benchmarks, which omit realistic radiative physics and other effects. We move on to production-grade tests that connect directly to observational data, which is done by incorporating a realistic neutron-star atmosphere, ISM absorption, and NICER’s instrument response. In this section, we compare the results of the CPU and GPU versions of the codebase. At previous section, we are not discussing the computation performance specifically, because the most time-consuming part of this PPM is the realistic atmosphere model table interpolation, so vast computation improvement is based on improving this atmosphere treatment strategy. Here in this section, we will compare both the modeling accuracy and the computation performance, using the benchmarks provided in \citet{Choudhury_2024_waveform}. As we discuss in the previous section, we are using different strategy on surface discretizations, instead of using the discretization of only the hotspot region, we consider the whole star surface discretization as default option, so when comparing the modeling accuracy of the surface elements in the hotspot, we using the active grid number in the hotspot as the comparison. The goal of this section is to demonstrate that the extreme geometry tests of \citet{Choudhury_2024_waveform}, including cases that are challenging to model accurately at standard resolutions in existing codes such as \texttt{X-PSI} and \texttt{IM}, can be computed at our standard setup with discrepancies well below the Poisson fluctuation level, while achieving $10^4$--$10^5\times$ speedups relative to the Ultra-resolution \texttt{X-PSI} runs. In this section, CPU tests were performed on an Intel i7-13700KF with single core and GPU tests on an NVIDIA GeForce RTX 4080 (16 GB).

Following \citet{Choudhury_2024_waveform}, we adopt the same five ``extreme" waveform tests to compare PPM treatments. Four tests use a single-temperature hot spot. All shapes are built only from circular components, either {emitting} or {masking}, because some code in that test limited to circular primitives. Masking components simply hide the flux from the region they cover. 
We excluded the double-temperature test case from \citet{Choudhury_2024_waveform} because the compactness parameter was set to an extreme value, rendering the emission spot always visible through gravitational light bending and permitting multiple imaging--an effect not treated by the our present code version.
Complete parameter setup and the exact geometries (rings, crescents, and the bithermal configuration) are tabulated and illustrated in \citet[][Table~1 and Figure~1]{Choudhury_2024_waveform}.

All tests share the same non-hotspot settings: distance $D=150\ \mathrm{pc}$; observer inclination $i=2.391~\mathrm{rad}$ measured from the rotational north pole (a viewing geometry under which the near pole remains visible across rotational phase), spin frequency at infinity $\nu=300~\mathrm{Hz}$, and neutral hydrogen column density $N_{\mathrm H}=2\times10^{19}\ \mathrm{cm^{-2}}$, and exposure time of $10^6$ s. The inclination is chosen to mirror the viewing of a well-studied nearby millisecond pulsar PSR~J0437$-$4715, and the spin frequency is set at $300$~Hz because the OS approximation is accurate to $\lesssim 0.1\%$ at and below this rate \citep{Bogdanov_19}, while discrepancies grow to a few percent by $\sim 600$~Hz \citep{Cadeau_2007, jakab2025gravitationalredshiftrapidlyrotating}. 

The four single-temperature tests use $M=1.4\,M_\odot$ and equatorial radius $R_{\rm e}=12.0~\mathrm{km}$. Within the single-temperature set, two tests form nonconcentric {rings} by nesting a mask inside an emitter (with sizes varied to stress overlap logic and discretization), and two form elongated {crescents} by letting a mask protrude into an emitter to accentuate edge effects.

Atmosphere setup is uniform across tests. Each emitting component is modeled as a geometrically thin, fully ionized hydrogen atmosphere; specific intensities are drawn from \texttt{NSX-H} lookup tables \citep{Ho_2001}. The resulting spectra are then attenuated using the same $N_{\mathrm H}$ above. For any instrument-response handling and the full per-component quantities, we refer the reader to \citet{Choudhury_2024_waveform}, whose setup we follow here.

\subsection{Resolution tiers setup}

We evaluate two resolution tiers designed for distinct purposes: (i) a {standard} tier that is sufficiently accurate for routine inference at practical cost and (ii) a {high-resolution} tier intended to match the fidelity of the ``Ultra-style" reference settings in \citep{Choudhury_2024_waveform}. To isolate the effect of surface discretization, we vary only the active number of surface patches within each hotspot. All other precision controlled parameters are held fixed at high values comparable to those used in rigorous code-verification runs.
This ensures that numerical errors associated with phase and energy discretization, along with per-patch ray statistics, are negligible. The remaining lever on accuracy and speed is the surface discretization.

Table~\ref{tab:tiers} summarizes the two tiers detailed resolution setup. The ``high-resolution'' tier uses an active $256\times256$ surface tiling per hotspot, which is {comparable} to the surface discretization commonly used in ``Ultra-level" resolution tier (e.g., in \texttt{X-PSI}). The ``standard'' tier uses $128\times128$ patches per hotspot and is intended to keep numerical systematics below typical Poisson counting uncertainties for representative NICER-like exposures, while offering a substantial speedup. Because our codebase differs in architecture and algorithms from other frameworks, a one-to-one mapping of resolution settings is not straightforward; by “comparable” we mean that our tiers achieve similar accuracy to the production-grade benchmarks reported by \citet{Choudhury_2024_waveform}.

\begin{table*}
\centering
\caption{Numerical settings for the two resolution tiers in this work. }
\label{tab:tiers}
\begin{tabular}{ccccccc}
\toprule
{Resolution Tier} & $N_{\rm active}$ & $N_{u}$ & $N_{c_\psi}$ & $N_{c_\alpha}$& $N_{\rm fp}$ & $N_{E_o}$ \\
\hline
std-resolution  & $128\times128$ &256&512&512& 192 & 128 \\
high-resolution & $256\times256$ &256&512&512& 192 & 128 \\
\hline\hline
\end{tabular}
\tablefoot{
Only the count of active surface patch $N_{\rm active}$ varies across tiers; all other precision parameters (the number of discretized compactness in lensing table $N_{u}$, the number of discretized $\cos\psi$ in lensing table $N_{c_\psi}$, the number of discretized $\cos\alpha$ in lensing table $N_{c_\alpha}$, the number of fine phase $N_{\rm fp}$, the number of down-sampling observed photon energy grid $N_{E_o}$) are held fixed.
}
\end{table*}

\begin{figure*}
    \centering
    \includegraphics[width=1.0\linewidth]{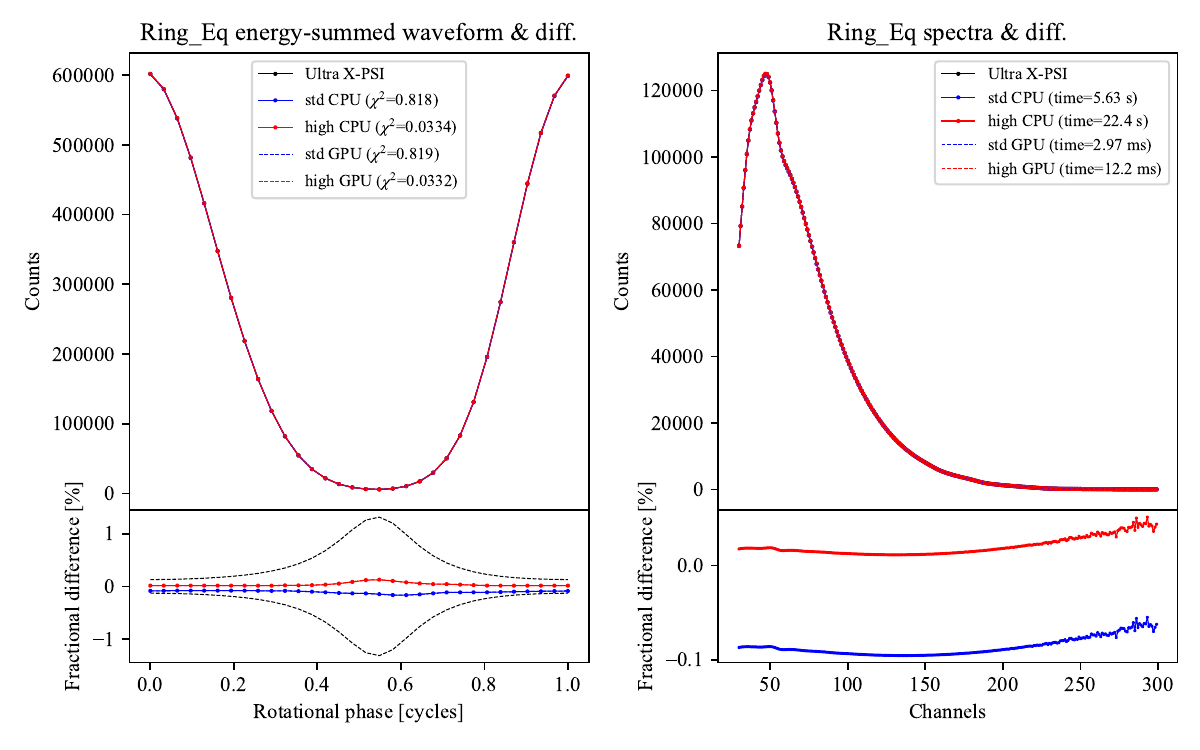}
    \caption{Energy-summed waveforms (left) and phase-summed spectra (right) for the \texttt{Ring-Eq} case. The bottom subpanels show the fractional difference (in \%) of each configuration relative to the X-PSI Ultra-resolution benchmark (black dotted). CPU results are plotted with solid lines: standard resolution (blue solid) and high resolution (red solid). GPU results are plotted with dashed lines: standard resolution (blue dashed) and high resolution (red dashed). In the left panel, the legend reports $\chi^2$ as defined in Equation~(\ref{chi2_def}), computed jointly over energy and phase channels as a quantitative measure of the discrepancy between full energy-resolved pulse profile. In the right panel, the legend lists the runtime for each configuration. Gray dashed lines in the bottom subpanels indicate the expected Poisson fluctuations of the Ultra-resolution \texttt{X-PSI} benchmark.}
    \label{fig:ring_eq}
\end{figure*}

\begin{figure*}
    \centering
    \includegraphics[width=1.0\linewidth]{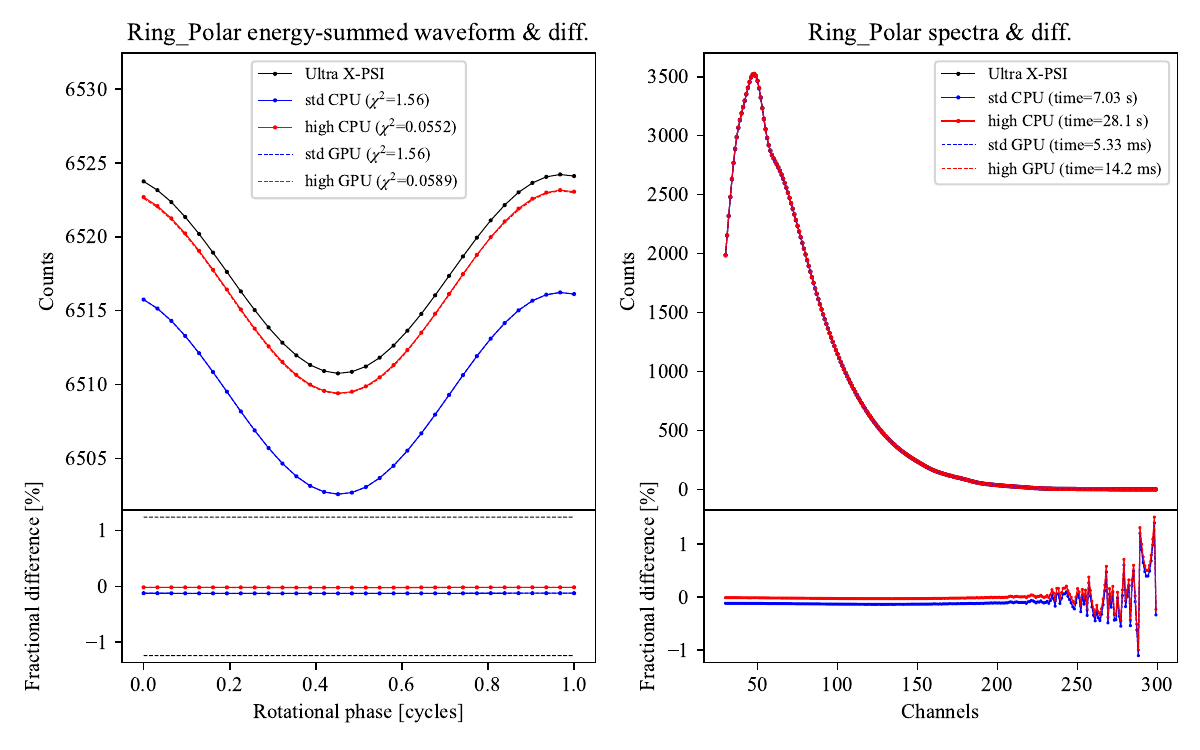}
    \caption{Energy-summed waveforms (left) and phase-summed spectra (right) for the \texttt{Ring-Polar} case. The bottom subpanels show the fractional difference (in \%) of each configuration relative to the \texttt{X-PSI} Ultra-resolution benchmark (black dotted). CPU results are plotted with solid lines: standard resolution (blue solid) and high resolution (red solid). GPU results are plotted with dashed lines: standard resolution (blue dashed) and high resolution (red dashed). In the left panel, the legend reports $\chi^2$ as defined in Equation~(\ref{chi2_def}) computed jointly over energy and phase channels as a quantitative measure of the discrepancy between full energy-resolved pulse profile. In the right panel, the legend lists the runtime for each configuration. Gray dashed lines in the bottom subpanels indicate the expected Poisson fluctuations of the \texttt{X-PSI} Ultra-resolution benchmark.}
    \label{fig:ring_polar}
\end{figure*}

\begin{figure*}
    \centering
    \includegraphics[width=1.0\linewidth]{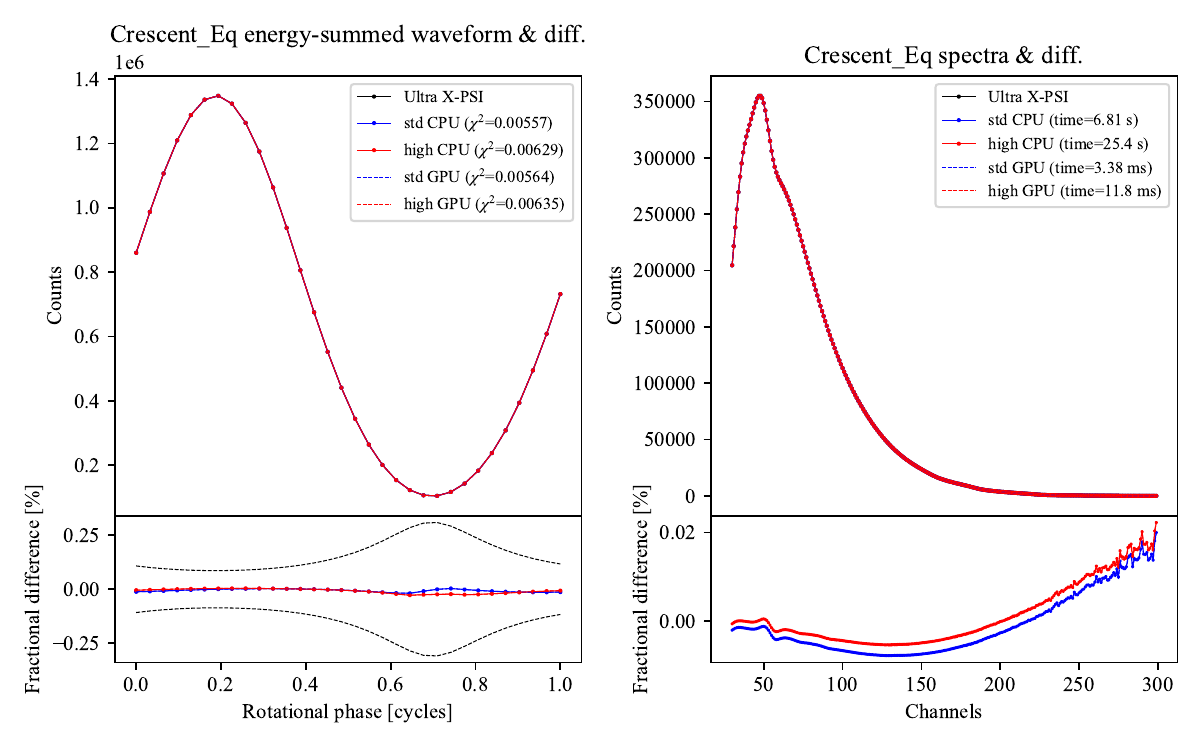}
    \caption{Energy-summed waveforms (left) and phase-summed spectra (right) for the \texttt{Ring-Polar} case. The bottom subpanels show the fractional difference (in \%) of each configuration relative to the \texttt{X-PSI} Ultra-resolution benchmark (black dotted). CPU results are plotted with solid lines: standard resolution (blue solid) and high resolution (red solid). GPU results are plotted with dashed lines: standard resolution (blue dashed) and high resolution (red dashed). In the left panel, the legend reports $\chi^2$ as defined in Equation~(\ref{chi2_def}) computed jointly over energy and phase channels as a quantitative measure of the discrepancy between full energy-resolved pulse profile. In the right panel, the legend lists the runtime for each configuration. Gray dashed lines in the bottom subpanels indicate the expected Poisson fluctuations of the \texttt{X-PSI} Ultra-resolution benchmark.}
    \label{fig:Crescent_Eq}
\end{figure*}

\begin{figure*}
    \centering
    \includegraphics[width=1.0\linewidth]{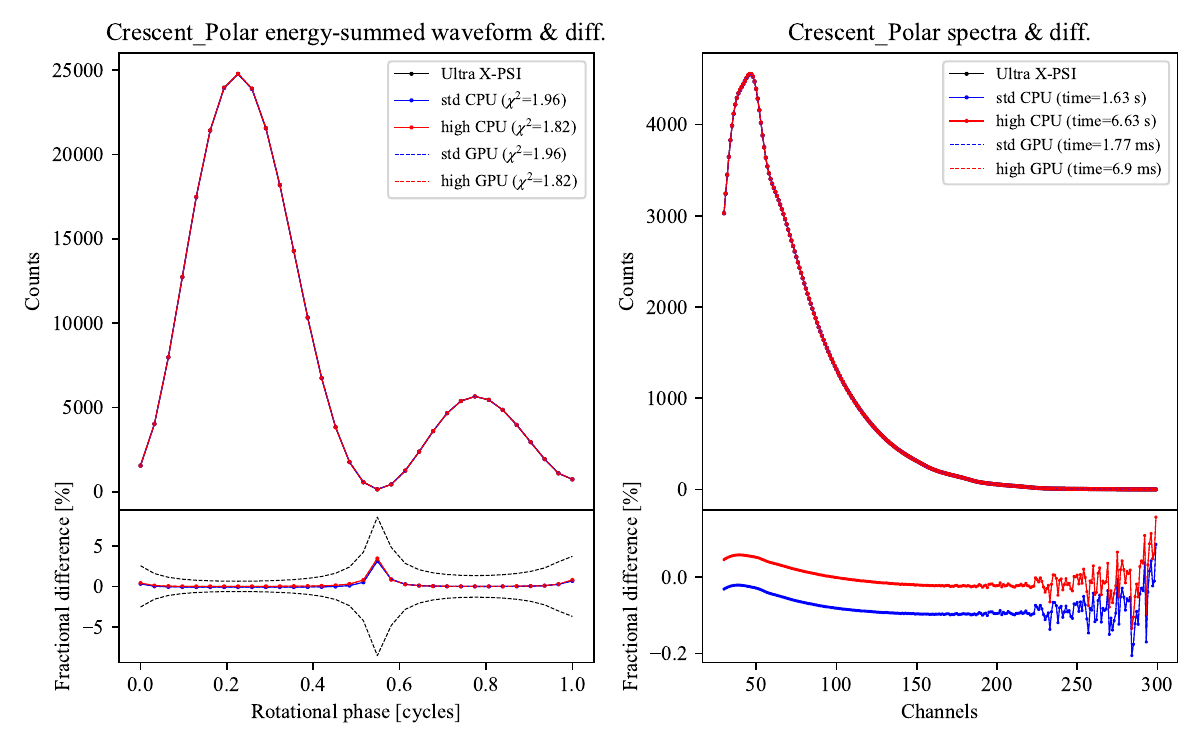}
    \caption{Energy-summed waveforms (left) and phase-summed spectra (right) for the \texttt{Ring-Polar} case. The bottom subpanels show the fractional difference (in \%) of each configuration relative to the \texttt{X-PSI} Ultra-resolution benchmark (black dotted). CPU results are plotted with solid lines: standard resolution (blue solid) and high resolution (red solid). GPU results are plotted with dashed lines: standard resolution (blue dashed) and high resolution (red dashed). In the left panel, the legend reports $\chi^2$ as defined in Equation~(\ref{chi2_def}) computed jointly over energy and phase channels as a quantitative measure of the discrepancy between full energy-resolved pulse profile. In the right panel, the legend lists the runtime for each configuration. Gray dashed lines in the bottom subpanels indicate the expected Poisson fluctuations of the \texttt{X-PSI} Ultra-resolution benchmark.}
    \label{fig:Crescent_Polar}
\end{figure*}

In our test runs, both CPU and GPU implementations produce phase-resolved, energy-resolved waveforms under identical physics and numerical settings. For validation, we compare our waveforms to a trusted reference (eg. \texttt{X-PSI} Ultra-resolution setup in \citet{Choudhury_2024_waveform}) and require fractional residuals $\lesssim 10^{-3}$ (i.e., $\lesssim 0.1\%$) over phases where the flux is non-negligible. Small, localized deviations may occur near the photon number extremes. These phases carry very low counts and have negligible impact on likelihood evaluations used in pulse profile inference. 

The reason we need to compare our CPU and GPU version code internally is because porting a ray-tracing pipeline from CPU to GPU is nontrivial. Memory layouts, parallel reduction order, and instruction-level differences can lead to small floating-point discrepancies even for the same numerical algorithm. While CPU implementations can reach either tier, end-to-end, high-resolution scans of the full parameter space are typically {computationally prohibitive} within reasonable computation time, whereas the GPU implementation enables production-grade inference at the high-resolution setting. The standard tier of GPU setup is designed to be extremely fast on GPU but already become untractable on CPU even with taking advantage of modern C++ coding, while keeping numerical systematics subdominant to Poisson fluctuation for all these extreme hot-spot setup. 

To demonstrate the robustness of our GPU code on the whole possible hotspot geometry parameter space, in the following subsections, we evaluate four geometries that stress the ray-tracing and integration pipeline: \texttt{Ring-Eq}, \texttt{Ring-Polar}, \texttt{Crescent-Eq}, and \texttt{Crescent-Polar}. Their precise definitions (spot shapes, locations, and masks) follow \citet[][Table~1 and Figure~1]{Choudhury_2024_waveform}. Each case is discussed in turn using identical numerical algorithm and precision-controlled parameters so that differences reflect only hotspot geometry or resolution. 
To ensure full transparency and reproducibility, we have released the source code at GitHub repository\footnote{\url{https://github.com/zhoutz/gpu_ppm}}. A versioned snapshot, including all scripts required to regenerate the figures for both CPU and GPU implementations at both resolution tiers, has been archived on Zenodo\footnote{\url{https://doi.org/10.5281/zenodo.17315493}}.

\subsection{Ring-Eq case}

The \texttt{Ring-Eq} configuration is a ring-shaped hotspot centered on the stellar equator. It typically produces a single predominant pulse per rotation. Figure~\ref{fig:ring_eq} compares energy-summed waveforms and phase-summed spectra across our two resolution tiers (standard and high) and across CPU and GPU backends, benchmarked against the \texttt{X-PSI} Ultra-resolution reference used in \citet{Choudhury_2024_waveform}. The \texttt{X-PSI} Ultra-resolution configuration serves here (and throughout) as a high-fidelity standard for extreme geometries.

To quantify the agreement, we can compute a joint, 2D $\chi^2$ statistic,
\begin{equation}
\chi^2 = K \sum_{i,j}\,\frac{\big(m_{ij}-d_{ij}\big)^2}{m_{ij}}\,,
\label{chi2_def}
\end{equation}
where $m_{ij}$ and $d_{ij}$ denote model and reference counts in energy bin $i$ and phase bin $j$, respectively. The normalization factor $K$ is defined as
\begin{equation}
K = \frac{10^6}{\sum_{i,j} m_{ij}}\,.
\end{equation}
The normalization constant is included to enable fair comparisons between different waveforms. Without it, Equation~(\ref{chi2_def}) implies that the overall $\chi^2$ would scale with the total photon counts. Accordingly, we normalize all computations to a fixed total of $10^6$ counts, as suggested in \citet{Choudhury_2024_waveform}. Under Poisson statistics and in the large-count limit, $-2\,\Delta\ln L \approx \chi^2$, so maintaining small $\chi^2$ controls the corresponding change in log-likelihood ($\Delta\ln L \approx -\tfrac{1}{2}\chi^2$). Our goal is to demonstrate that both CPU and GPU implementations meet production-grade accuracy for pulse profile inference.

The energy-summed waveforms from all four runs (standard+high resolution $\times$ CPU+GPU) agree with the benchmark to within the expected Poisson fluctuations. Differences cluster by resolution rather than by backend: increasing the surface-tiling density (from $128\times128$ to $256\times256$ patches per hotspot) uniformly reduces fractional residuals, while switching from CPU to GPU at fixed resolution has no visible effect on accuracy. Small deviations concentrate near flux minima (when the ring rotates partially out of view), where counts are lowest. These phases carry negligible statistical weight in likelihood evaluations.

The phase-summed spectra exhibit the same pattern. Curves at a given resolution overlap closely regardless of backend, and the high-resolution tier yields uniformly smaller absolute fractional differences. This behavior is consistent with surface discretization being the dominant residual error once the per-patch rays, phase sampling, and energy sampling are held fixed at high settings.

Runtime measurements are shown in the legend of Figure~\ref{fig:ring_eq}. Within our codebase, increasing the surface-patch count by a factor of four (standard $\rightarrow$ high resolution) scales wall time by approximately the same factor (e.g., $5.63$\,s $\rightarrow$ $22.4$\,s on CPU), as expected for a largely linear compute-to-patch workload. Even with careful modern C++ implementation, exhaustive parameter-space exploration at high resolution is computationally prohibitive on CPU, whereas the GPU implementation achieves $\sim 2000\times$ speedups over our CPU runs at identical settings, enabling production-grade inference even under the high-resolution setup. For context, the quoted \texttt{X-PSI} Ultra-resolution runtime for comparable resolution is $\sim 1$--$5$ minutes; our GPU timings are $4$--$5$ orders of magnitude faster while maintaining agreement at the $\chi^2\sim 10^{-2}$ level.

\subsection{Ring-Polar case}

The \texttt{Ring-Polar} configuration is a small ring-shaped hotspot centered near the rotational pole. Because the emitting area is small, a spot at the same surface temperature produces fewer total counts than larger hotspot. Moreover, for typical neutron-star compactness (e.g., $M\approx1.4\,M_\odot$, $R_{\rm e}\sim12$\,km), gravitational light bending keeps most of a polar ring at least partially in view over the rotation cycle for a wide range of observer inclinations, so the modulation amplitude is modest rather than strongly pulsed.

Figure~\ref{fig:ring_polar} compares energy-summed waveforms and phase-summed spectra across our two resolution tiers (standard and high) and across CPU and GPU backends, using the \texttt{X-PSI} Ultra-resolution configuration as the high-fidelity reference. Agreement is quantified by the joint, 2D statistic $\chi^2$. In our runs, energy-summed waveforms from all configurations are consistent with the benchmark at the Poisson level: the {standard} tier yields $\chi^2 \approx 1.5$ for both CPU and GPU, while the {high-resolution} tier reduces this to $\chi^2 \approx 0.05$. As in the other cases, differences group by resolution rather than by backend, indicating that surface discretization, not CPU vs.\ GPU, is the dominant factor governing residuals once other precision settings are held fixed.

The phase-summed spectra show the same pattern. At energies where the counts are highest, curves overlap closely; in the highest-energy channels (e.g., channel indices $\gtrsim 250$, where counts are lowest), fractional residuals can reach the $\sim 1$--$1.5\%$ level without changing the total $\chi^2$. This behavior is expected: low-count bins carry little statistical weight, so localized percent-level differences there have negligible impact on the likelihood computation.

Within a given backend, as in the \texttt{Ring-Eq} case, increasing the surface-patch count from $128\times128$ to $256\times256$ (a $\times4$ increase) scales the \texttt{Ring-Polar} runtime by approximately the same factor, as expected for a patch-parallel workload. Absolute runtime for the \texttt{Ring-Polar} case is comparable to, or slightly longer than, that for the equatorial ring. While the CPU implementation meets the accuracy target, exhaustive parameter scans at either resolution tier are computationally prohibitive on CPU within practical budgets. The GPU implementation achieves $\sim 10^3\times$ speedups at identical settings, enabling production-grade Bayesian inference at very high fidelity.

\subsection{Crescent-Eq case}

In the \texttt{Crescent-Eq} configuration, the hotspot is an elongated crescent located near the stellar equator. Because the crescent spans a substantial azimuthal extent, it remains visible over a large fraction of the rotational phase and yields comparatively high counts. As noted by \citet{Choudhury_2024_waveform}, this geometry can be slightly more sensitive to code implementations and resolution choices, with percent-level differences sometimes appearing at phases where the flux approaches zero; such localized discrepancies are expected when counts are minimal and carry little statistical weight.

In our experiments (Figure~\ref{fig:Crescent_Eq}), the energy-summed waveforms agree with the \texttt{X-PSI} Ultra-resolution benchmark at the $\lesssim 0.1\%$ level across phase, and increasing the surface-patch density from the standard tier ($128\times128$ patches per hotspot) to the high-resolution tier ($256\times256$) produces no visually discernible change. The joint 2D statistic $\chi^2$ defined earlier likewise remains very small: at the standard tier we obtain $\chi^2 \approx 0.005$, while at the high-resolution tier we find $\chi^2 \approx 0.006$. Differences of this magnitude are consistent with floating-point and discretization effects rather than systematic modeling errors; at such small values, monotonic improvements in $\chi^2$ with resolution are not guaranteed.

The phase-summed spectra show a similar pattern. Curves overlap closely in energy ranges with substantial counts, while the highest-energy bins can exhibit slightly larger fractional residuals without materially affecting the total $\chi^2$. Small backend-dependent trends (CPU vs.\ GPU) in these highest-energy bins are consistent with floating-point accumulation effects. At fixed resolution, results group by resolution rather than by hardware.

We attribute the robustness in this geometry, in part, to our full-surface discretization strategy: the {entire} stellar surface is finely meshed, and the crescent is selected by masking within that mesh. This approach represents sharp crescent edges cleanly across phase, reducing edge-alignment artifacts that can arise when only the hotspot boundary is discretized. While alternative pipelines employ different but valid tiling schemes \citep[e.g.,][]{Bogdanov_19,Bogdanov_21}, our results indicate that a full-surface mesh provides stable accuracy for elongated, edge-dominated spots.

Finally, the runtime remains highly favorable. On the GPU, the standard tier computes a single energy-resolved profile in $\sim 3.4$\,ms, enabling production-grade scans. On the CPU, the same configuration requires $\sim 7.16$\,s at the standard tier and $\sim 27$\,s at the high-resolution tier. Thus, although the CPU implementation attains the desired accuracy, only the GPU implementation provides the required speedup to support large-scale Bayesian inference. 

The crescent geometry is particularly relevant for future studies incorporating higher-order magnetic multipoles, which can naturally yield noncircular, elongated hotspots. For example, \citet{Huang:2025_hotspot} show that, in an off-centered dipole magnetosphere, incorporating a volumetric return-current region modifies the polar-cap temperature map, adding a crescent-like, lower-temperature component. These effects are modest for a centered (aligned) dipole but become more pronounced as the dipole offset increases. Thus, testing this crescent configuration under different placements is extremely valuable for future physics-motivated, hotspot-based PPM. 

\subsection{Crescent-Polar case}
\label{subsec:crescent_polar}
In the \texttt{Crescent-Polar} configuration, we place the same elongated hotspot near the rotational north pole and adopt a near-limb viewing geometry, such that only a small portion of the emitting region is directly visible at any given phase. In this edge-dominated regime, most detected photons originate from a narrow strip along the spot boundary where projected area, light bending, and atmospheric beaming combine to produce a steep intensity gradient. Consequently, the energy-summed waveform becomes highly sensitive to numerical resolution specifically at the spot edge. Across different codes, and even within the same code at different resolution settings, this geometry can produce visibly different waveforms. Although the largest fractional differences tend to occur in low-count bins, the associated mismatch metric, $\chi^2$ (as defined in Equation~(1) of \citet{Choudhury_2024_waveform} and here computed jointly over energy and phase channels), can still become large enough to change the likelihood by a factor $\exp(-\chi^2/2)$, potentially biasing inference if not controlled.

Figure~\ref{fig:Crescent_Polar} shows that our implementation closely matches the benchmark in this challenging configuration. Using both the CPU and GPU pipelines at the standard resolution tier yields $\chi^2 \approx 1.96$. In typical PPM analyses, a forward-model mismatch of order unity is negligible relative to the $\chi^2$ variations that drive parameter updates and therefore has no visible impact on the inference. At the high-resolution tier, the discrepancy remains at $\chi^2 \simeq 1.82$, indicating only marginal improvement in precision. The right panel shows that the spectral fractional differences decrease at high resolution, this is the primary source of the modest improvement from the standard to the high-resolution tier. By contrast, the left panel exhibits a persistent pattern in the fractional differences that generally stays below the Poisson fluctuation level; however, near half phase (the trough between the two peaks), the residuals reach up to $\sim 5\%$. This phase interval corresponds to when the crescent-shaped polar hotspot becomes faint and approaches the limb ($\mu \to 0$). As discussed in Section~\ref{sec4:atmosphere_interpolation}, cubic-interpolation overshoot in this regime, followed by the non-negativity clipping safeguard in \texttt{X-PSI}, leads to an underestimation of the local photon counts. While the clipping prevents the systematic from exceeding the Poisson fluctuation level, it nevertheless produces a real, phase-localized bias that is evident here and results in an overall underestimation of the counts. Comparing with Figure~7 of \citet{Choudhury_2024_waveform}, the \texttt{IM} group’s results oscillate between under- and overestimating the photon counts relative to the \texttt{X-PSI} benchmark, suggesting that their atmosphere-model interpolation is sensitive to interpolation choices. In contrast, the \texttt{Alberta} codebase, which uses linear interpolation for the atmosphere model, is more robust at boundaries and does not suffer from overshoot-induced underestimation. Accordingly, it yields consistently higher counts at the limb. These comparisons reinforce that careful treatment of atmosphere-table interpolation is essential for reliable pulse profile computations. Consistent with our other tests, the phase-summed spectra indicate that the residual fractional differences are dominated by high-energy channels with very low counts.

From a performance standpoint, our code remains highly efficient in this case. On the CPU, the standard setup runs in $\sim 1.63$\,s and the high-resolution setup in $\sim 6.63$\,s. On the GPU, the corresponding run times are $\sim 1.77$\,ms and $\sim 6.9$\,ms, respectively, yielding speedups of order $10^3$ relative to our own CPU implementation for both resolution tiers. These gains are achieved without emulators or machine-learning surrogates; we use high-fidelity forward calculations with algorithmic refinements executed on the GPU.

This case underscores an important methodological point. Because edge-on, elongated spots can exist in physically motivated scenarios (e.g., modestly off-centered dipoles can place volumetric hotspot regions near a rotational pole; see Figures~3a--b of \citet{Huang:2025_hotspot}), there is no compelling physical argument to {a priori} forbid such geometries. Reliable inference therefore requires numerical treatments that remain accurate in edge-dominated regimes. By breaking the traditional speed-versus-accuracy trade-off through a GPU implementation, our code provides one practical pathway to control $\chi^2$ to the $\sim 1$ level even in this extreme configuration, while simultaneously achieving $\sim 10^3$--$10^4$ speedups.

Our CPU implementation uses IEEE~754 double-precision floating-point numbers (64-bit, FP64), whereas the current GPU implementation on an RTX~4080 employs single precision (32-bit, FP32) for all computations. Because our forward model involves no long time integration, cumulative round-off error is limited; moreover, we adopt numerically stable formulations to control FP32 error. In the current benchmarks at matched resolution, the FP32 GPU results are indistinguishable from the FP64 CPU results, indicating that FP32 is adequate for our present configuration. From an engineering perspective, enabling FP64 in our GPU codebase is straightforward. Users who run this code on scientific GPU clusters may optionally switch to FP64 to improve numerical precision.

\section{Discussion and conclusion}
\label{sec6:discussion}

This work presents a GPU-accelerated X-ray pulse profile modeling (PPM) codebase that directly addresses the typical accuracy-speed trade-off in current analyses. We first outline the theoretical framework of pulse profile forward modeling, rigorously verifying and deriving all formulas essential to PPM forward computation (with more details in Appendices~\ref{AppendixA:doppler}-\ref{AppendixC:flux_formula}). We then provide a detailed computational recipe for both CPU and GPU implementations and clearly document the porting of the CPU code to the GPU to maximize reproducibility.

We verified our implementation by first reproducing the theoretical light-curve benchmarks of \citet{Bogdanov_19}, demonstrating accurate computation of light curves for specified spacetime and hotspot geometries under the oblate-Schwarzschild (OS) approximation. The Schwarzschild\,+\,Doppler (S+D) approximation is also supported, although OS remains our default production setting.

We further examined a commonly ignored source of sytematic error that arises when moving from theoretical benchmarks to models with realistic atmospheres. Because of the current structure of the widely used NSX model, several codebases interpolated within these precomputed neutron-star atmosphere lookup tables using high-order polynomial interpolation. Using two representative viewing-geometry test cases, we demonstrate that cubic Lagrange interpolation can produce significant overshoot, yielding unphysical negative specific intensities. One approach to mitigation is to clip these negative values to zero. However, in the test cases presented here, both strategies (cubic Lagrange interpolation with or without clipping) can lead to underestimation of the total photon flux at many rotational phases, particularly in the soft X-ray band, depending on implementation. Preliminary X-PSI tests indicate smaller negative flux values (T. Salmi, priv. comm.) and the phase–energy locations of any deficits may differ across implementations.

Moreover, these new test cases are neither extreme nor exotic. They plausibly represent the emission geometry of a southern polar cap arising from either a canonical centered dipole or an off-centered dipole configuration. The region of parameter space that exhibits this overshoot behavior is potentially broad; modest variations of the pulse profile model parameters readily produce similar behavior across different settings. In fact, for many common viewing and hotspot geometries, rotation will bring portions of the hotspot into and out of the line of sight near the limb, where these interpolation artifacts are most likely to arise. The interpolation artifact identified here suggests that previously reported PPM inferences that relied on high-order polynomial interpolation without boundary-aware treatment could contain an uncontrollable systematic error. Reexamining those analyses with a boundary-aware interpolation scheme, like the one proposed in this work, would be extremely valuable for assessing the impact on inferred parameters, especially $M$ and $R$.

To mitigate interpolation artifacts, we adopted a mixed-order interpolation scheme. Within the interior of the atmospheric lookup tables, we used cubic Lagrange interpolation; but at the grazing-emission boundary, we reverted to linear interpolation to suppress overshoot. We also allow the interpolation order to differ across table dimensions, mixing linear and cubic along different axes to reduce computational cost while maintaining accuracy. The optimal configuration is table- and physics-dependent; constructing denser atmospheric grids would enable a more definitive validation of the interpolation scheme. A comprehensive evaluation is deferred to future work.

Building on these theoretical checks and the careful treatment of the atmosphere model, we turned to production-grade benchmarks, which include all of the necessary ingredients for PPM forward computation. We reproduced the suite of benchmarks in \citet{Choudhury_2024_waveform}, incorporating a realistic atmospheric treatment, interstellar absorption, and the NICER instrument response. We implemented both CPU and GPU codes and examined two resolution configurations that differ primarily in the number of surface grid elements within the hotspot. Holding all other setup choices fixed allowed us to isolate how discretization influences runtime scaling while maintaining comparable physical fidelity.

The observed GPU performance indicates clear implications for inference, enabling higher-resolution forward models, broader parameter exploration, and more robust convergence diagnostics within modest computational budgets. With a modern C++ implementation, CPU runtimes at these extremely high resolutions, ranging from several seconds to tens of seconds per profile, remain prohibitive for full Bayesian analyses and robustness testing. In contrast, the GPU implementation presented here reduces per-profile computation time to the millisecond regime for the standard resolution and to $\sim 10$~ms for the extremely high-resolution setting on a single RTX~4080 GPU. Because fully converged posterior exploration can require $10^7$-$10^8$ likelihood evaluations (depending on the problem's dimensionality), these timings make end-to-end inference feasible on a single GPU or modest GPU clusters. 

Crucially, the speed improvement does not compromise accuracy. At the fine resolutions adopted here, our results closely match published benchmarks while retaining millisecond-scale runtimes. Even for geometries previously viewed as resolution-sensitive, our high-resolution GPU runs reproduce reference benchmarks, effectively removing the hard accuracy-speed bottleneck.

Beyond the common single- or two-temperature prescriptions used in many current inferences, our framework accepts arbitrary, user-specified global temperature maps on the stellar surface. This capability enables inference based on physics-motivated hotspots, such as those predicted by off-centered dipole or more general multipolar magnetic configurations \citep[e.g.,][]{Huang:2025_hotspot,Gralla2017,Lockhart2019}, and naturally extends to accretion-powered millisecond pulsars, where global temperature distributions are more appropriate than bounded caps \citep{Das_2022,Das_2025}. Because purely geometric hotspot choices can introduce degeneracies and ad hoc modeling systematics, we advocate reducing arbitrary geometry in favor of models tied to the magnetospheric structure of the neutron star (e.g., controlled multipolar expansions) and potentially enabling multiwavelength constraints. Related efforts \citep[e.g.,][]{Kalapotharakos_2021,Olmschenk:2025eie,Petri:2025nhu} typically assume uniform-temperature caps; observational and theoretical indications suggest, however, that polar-cap regions are structured, with nonuniform temperatures and subregions dominating the X-ray emission \citep{Spitkovsky_2006,Bai_2010,Gralla2017,Lockhart2019}. Our code is designed to accommodate such more complex hotspot models.

While the present work emphasizes fast and accurate forward modeling, the inference layer is a natural next step. Our GPU optimization extends beyond waveform synthesis to the {entire} likelihood evaluation: phase- and energy-resolved model counts, background treatment, and Poisson log-likelihood accumulation run on GPU and parallelize naturally across energies, phase bins on different parameters. We support two complementary paths for sampling: (i) a thin Python/C++ interface that interoperates with established CPU-side samplers; for instance, \textsc{MultiNest} \citep{multinest}, \textsc{UltraNest} \citep{UltraNest}, and \textsc{Dynesty} \citep{Dynesty}. We also kept the likelihood kernel resident on the GPU to minimize host--device overhead; and (ii) a fully GPU-resident parallel-tempered MCMC (PT-MCMC); and, resources permitting, a GPU-friendly nested-sampling engine) to handle multimodal posteriors and strong parameter correlations. Both modes can scale to multiple GPUs by sharding chains or live points across GPUs and batching likelihood calls within each device. Using this architecture together with physics-motivated temperature maps, we will reanalyze NICER sources, systematically varying samplers and settings to assess robustness and to deliver stable, reproducible constraints on neutron-star masses and radii.

Before proceeding to inference, we note an additional physical effect: multiple imaging due to strong-field gravitational light bending. For sufficiently compact stars, approximately when $GM/(Rc^{2}) \gtrsim 0.28$, photon trajectories from a given surface element can reach the observer along multiple paths, yielding secondary images. Our current framework is not yet eqiupped to model these higher-order images and, thus, implementing this capability for production-grade inference is deferred to future work. For context, current NICER constraints for PSR~J0740+6620 imply a central-value compactness likely below this threshold, although portions of the allowed parameter space may reach it; in those more compact configurations, this multiple-imaging effect could become relevant.

In summary, the codebase presented here provides a modern computational framework for PPM: it achieves production-grade accuracy at millisecond-scale runtimes. Our results are aligned with the majority of previous community tests and we propose two new atmosphere-interpolation tests. Our work substantially breaks the accuracy-speed bottleneck of Bayesian PPM studies. In parallel with advancing reanalyses of existing observational sources with physics-motivated hotspot models, we invite the community to use our GPU PPM codebase to reexamine the robustness of previous studies under diverse inference setups. For transparency and pedagogical reasons, we also provide a CPU reference implementation with a clear structure and high readability. Taken together, these efforts advance pulse profile modeling toward a more efficient, reliable, and reproducible regime.

\section*{Data availability}
GPU-PPM Github repository: \url{https://github.com/zhoutz/gpu_ppm}. Zenodo repository of reproduction file of this work:  \url{https://doi.org/10.5281/zenodo.17315493}

\begin{acknowledgements}
H.C. acknowledges support from NSF Grant AST-2308111.  The authors extend gratitude to Sharon Morsink, Tuomo Salmi for providing valuable Benchmark, help on clarifying the theoretical details about PPM, Alexander Chen about the GPU implementation suggestions and H.C. would like to thank Anna Watts, Alexander Dittmann, Cole Miller, Roger Romani, Wenbin Lu, Roger Blanford for insightful discussion and comments. 
\end{acknowledgements}

\bibliographystyle{aa}
\bibliography{sample701}

\begin{appendix}

\section{Relativistic Doppler effect}
\label{AppendixA:doppler}
To account for the relativistic Doppler effect, we consider two inertial reference frames: the {lab} frame, in which the stellar surface element is in motion, and the {comoving} frame, in which the element is at rest. Throughout this section, we set light speed $c=1$. A prime ($'$) denotes quantities measured in the {comoving} frame.
The goal is to relate (i) the angle between the photon direction and the surface normal, and (ii) the solid-angle element, as seen in the two frames. These relations are what enter the specific-intensity transformations.

Suppose that, in the {lab} frame, a small emitting surface patch translates along the $+\hat y$ direction with speed $v$, while its outward surface normal is aligned with $\hat z$. A photon leaves the patch in direction $\hat k$. We project $\hat k$ onto the $x$-$z$ plane. Define $k_\perp$ in $x$-$z$ plane,
\begin{equation}
k_\perp = \hat{k}-\hat{k} \cdot \hat{y}, \quad
\hat{k}_\perp = \frac{k_\perp}{|k_\perp|}.
\end{equation}

We then parametrize the direction by a polar angle $\xi$ measured from $+\hat y$ direction and an azimuth $\phi$ measured in the $x$-$z$ plane. With this choice,
\begin{equation}
\hat{k} \cdot \hat{y} = \cos \xi,\quad
\hat{k}_\perp \cdot \hat{x} = \cos \phi,\quad
\hat{k}_\perp \cdot \hat{z} = \sin \phi,
\end{equation}
so that the unit propagation vector has the component form
\begin{equation}
\hat{k} = (\sin\xi \cos\phi, \cos\xi, \sin\xi\sin\phi).
\end{equation}
Along the null ray, the spatial coordinates advance at light speed in the direction $\hat k$, hence
\begin{equation}
x = t \sin\xi\cos\phi,\quad
y = t \cos\xi,\quad
z = t \sin\xi\sin\phi,
\end{equation}
We now boost to the {comoving} frame with a standard Lorentz transformation along $+\hat y$:
\begin{equation}
t' = \gamma(t-vy),\quad
y' = \gamma(y-vt),\quad
x' = x,\quad
z' = z,
\end{equation}
where the Lorentz factor is
\begin{equation}
\gamma = \frac{1}{\sqrt{1-v^2}} .
\end{equation}
Next, we introduce the angle $\sigma$ between the photon direction and the surface normal $\hat z$. This is the local emission angle relevant for atmosphere beaming, in our geometry,
\begin{equation}
\cos\sigma = \frac{z}{\sqrt{x^2+y^2+z^2}} = \sin\xi\sin\phi,
\end{equation}
and the same definition applied in the {comoving} frame gives $\sigma'$. Using the boosted coordinates and the null condition, we obtain
\begin{equation}
\begin{split}
\cos\sigma' &= \frac{z'}{\sqrt{x'^2+y'^2+z'^2}} \\
&= \frac{\sin\xi\sin\phi}{\sqrt{(\sin\xi\cos\phi)^2+[\gamma(y/t-v)]^2+(\sin\xi\sin\phi)^2}} \\
&= \frac{\cos\sigma}{\sqrt{\sin^2\xi+\gamma^2(\cos\xi-v)^2}} \\
&= \frac{\cos\sigma}{\gamma \sqrt{(1-v^2)\sin^2\xi+\cos^2\xi-2v\cos\xi+v^2}} \\
&= \frac{\cos\sigma}{\gamma \sqrt{1-2v\cos\xi+v^2\cos^2\xi}} \\
&= \frac{\cos\sigma}{\gamma (1-v\cos\xi)} \\
&= \delta \cos\sigma ,
\end{split}
\end{equation}
where we have identified the usual Doppler factor
\begin{equation}
\delta = \frac{1}{\gamma(1-v\cos\xi)} .
\end{equation}
This expression shows explicitly how the emission angle with respect to the surface normal is modulated by the boost.

We now turn to the transformation of the solid angle. With our spherical coordinates we define
\begin{equation}
d \Omega = d \cos\xi d \phi,\quad
d \Omega' = d \cos\xi' d \phi',
\end{equation}
Because the Lorentz boost is along $y$ and leaves $x$ and $z$ unchanged, the azimuth in the $x$-$z$ plane is unaffected:
\begin{equation}
\phi' = \arctan \left(\frac{z'}{x'}\right) = \arctan \left(\frac{z}{x}\right) = \phi ,
\end{equation}
so that
\begin{equation}
d \phi = d \phi'.
\end{equation}
The polar cosine with respect to the boost axis is, by definition,
\begin{equation}
\cos\xi = \frac{y}{\sqrt{x^2+y^2+z^2}} ,
\end{equation}
and its correspondence in {comoving} frame is
\begin{equation}
\cos\xi' = \frac{y'}{\sqrt{x'^2+y'^2+z'^2}}
= \frac{\gamma (y/t - v)}{\gamma (1-v \cos\xi)}
= \frac{\cos\xi - v}{1-v \cos\xi} .
\end{equation}
This is the standard formula for angles measured relative to the boost direction. Differentiating with respect to $\cos\xi'$ gives the Jacobian of the transformation:
\begin{equation}
\begin{aligned}
d\cos\xi'
&= \frac{
d\cos\xi(1-v\cos\xi) - (\cos\xi - v)\,(-v\, d\cos\xi)
}{
(1-v\cos\xi)^2
} \\
&= d\cos\xi \,\frac{1-v^2}{(1-v\cos\xi)^2} \\
&= \delta^2\, d\cos\xi \, .
\end{aligned}
\end{equation}
which immediately implies the solid-angle relation
\begin{equation}
d\Omega' = \delta^2 d\Omega.
\label{eq:dOmega}
\end{equation}

\section{Time dilation approximation}
\label{AppendixB:time_dilation}
In this section, we derive the mapping between emission time at a rotating stellar surface element and the reception time at a distant detector. The goal is to obtain a compact differential relation $dt_{\rm emit}\leftrightarrow dt_{\rm receive}$

Suppose a nongravitating spherical star which center locate at $(x,y,z)=(0,0,0)$ with mass $M$ and equatorial radius $R_{\rm e}$, a distant observer is located in the $+x$ direction.
The location of a small beaming patch on the star surface while rotating we denote as $\bm{r}(t_{\rm emit})$(all $t$ are measured in observer's frame).
A photon emit from $\bm{r}$, that finally hit the plane $x=D$, where $D$ is a large number represent the star-observer distance, will undergo 2 stages. 

The 1st stage is, from the photon been emitted, to it arrive the plane $x=R$, we use $\Delta t_1$ to denote the the time elapsed in stage 1,
\begin{equation}
t_{\rm hitR} = t_{\rm emit} + \Delta t_1.
\end{equation}
Here, $x=R$ is a fiducial plane placed beyond the stellar surface along $+\hat{x}$. 

The 2nd stage begins when photon hit $x=R$ and ends when photon hit plane $x=D$,
\begin{equation}
t_{\rm receive} = t_{\rm hitR} + \Delta t_2,\quad
\Delta t_2 = \frac{D-R}{c}.
\end{equation}
Because the ray is taken to propagate parallel to $+\hat x$ once it leaves the near zone, $\Delta t_2$ is a constant independent of emission phase. This cleanly separates phase-dependent (Stage 1) from phase-independent (Stage 2) contributions.

In the 1st stage, we consider 2 consecutive emitted photon from the same patch, photon A and photon B, with emitting time relation
\begin{equation}
t^B_{\rm emit} = t^A_{\rm emit} + dt_{\rm emit},
\end{equation}
since the star is rotating, these 2 photon are not emit at the same location, so we have $\Delta t^A_{1} \ne \Delta t^B_{1}$.
Since we are treating the nongravitating case, we ignore the gravitational light bending effect, we assume the photon go straight line towards $+x$ direction.
From geometry we can derive
\begin{equation}
\Delta t_1 = \frac{R-\hat{x}\cdot\bm{r}(t_{emit})}{c} ,
\end{equation}
so
\begin{equation}
\frac{d (\Delta t_1)}{d t_{emit}}
= \frac{-\hat{x}\cdot \dot{\bm{r}} (t_{emit})}{c}
= - \beta \cos\xi ,
\end{equation}
where $\cos\xi = \hat{x} \cdot \hat{\bm{v}}$ is the cosine of the angle between patch velocity and the photon emit direction, and
\begin{equation}
\beta = \frac{v}{c}=\frac{|\dot{\bm{r}}|}{c}.
\end{equation}

In stage 1, we get
\begin{equation}
\frac{d t_{\rm hitR}}{d t_{\rm emit}} = 1+\frac{d (\Delta t_1)}{d t_{\rm emit}}
= 1 - \beta \cos\xi ,
\end{equation}
which expresses the local time-compression/expansion between emission and crossing of the $x=R$ plane. 

In stage 2, $\Delta t_2$ is constant, so
\begin{equation}
\frac{d t_{\rm receive}}{d t_{\rm emit}} = \frac{d t_{\rm hitR}}{d t_{\rm emit}}
= 1 - \beta \cos\xi ,
\end{equation}
and the total mapping from emission to reception, in the nongravitating straight-line limit, is purely kinematic. When the surface moves toward the observer ($\cos\xi>0$) the received cadence is compressed, when it recedes, it is dilated.

When considering a Schwarzschild gravitating star, the surface of star and the observer have different gravitational potential, so we have to include the gravitational redshift effect, which is
\begin{equation}
\frac{d t_{\rm receive}}{d t_{\rm emit}} = 
\frac{1}{\sqrt{1-R_{\rm S}/R}} ,
\end{equation}
where $R_{\rm S}$ is the Schwarzschild radius associated with $M$ and $R$.

Combine both, we have
\begin{equation}
d t_{\rm emit} = \frac{\sqrt{1-R_{\rm S}/R}}{1 - \beta \cos\xi} d t_{\rm receive} ,
\label{eq:dt_emit_receive}
\end{equation}
which is the working relation used in the main text. The numerator represents gravitational time dilation, and the denominator represents kinematic compression/expansion from motion along the line of sight. In the limiting cases $\beta\to 0$ or $R_{\rm S}/R\to 0$, Equation~(\ref{eq:dt_emit_receive}) correctly reduces to purely gravitational or purely kinematic effects, respectively. If both vanish, $dt_{\rm emit}=dt_{\rm receive}$ as expected.

\section{Blackbody radiation and flux formula}
\label{AppendixC:flux_formula}

We begin with the Planck's law of blackbody radiation,
\begin{equation}
B(\nu, T) = \frac{2h\nu^3}{c^2} \frac{1}{e^{h\nu/kT} - 1},
\end{equation}
where $B(\nu, T)$ is the spectral radiance of a blackbody in thermal equilibrium at temperature $T$, with
unit power per (unit solid angle $\times$ unit area normal to the propagation $\times$ unit frequency).
Express it in more precise form,
\begin{equation}
\frac{EdN}{dtd\Omega\cos\sigma dSd\nu}=\frac{2h\nu^3}{c^2} \frac{1}{e^{h\nu/kT} - 1},
\end{equation}
where $E$ is the photon energy, $dN$ the photon count, $d\Omega$ the differential solid angle around the propagation direction, $\cos\sigma\, dS$ the projected surface area element, and $d\nu$ the frequency bin. Writing Planck’s law in this fully differential form will let us transform variables and frames cleanly.

The advantage of using this form is that we can easily change variables,
suppose we want to change $\nu$ to $E$ with
\begin{equation}
E=h \nu,\quad dE = h d\nu,
\end{equation}
we directly plug them in and get
\begin{equation}
\frac{EdN}{dtd\Omega\cos\sigma dSdE}=\frac{2E^3}{h^3 c^2} \frac{1}{e^{E/kT} - 1}.
\end{equation}
This recasts the blackbody intensity in per-energy units, which is convenient  for applications on detector responses that are tabulated versus $E$.

We use prime($'$) to denote quantities measured in the {comoving} frame of the emitting patch.
For a small patch on the star surface emitting blackbody radiation, we rewrite the above equation in {comoving} frame,
\begin{equation}
\frac{E'dN'}{dt'd\Omega'\cos\sigma' dS'dE'}=I_{bb} \equiv \frac{2E'^3}{h^3 c^2} \frac{1}{e^{E'/kT'} - 1}.
\end{equation}
Here $I_{bb}$ is the {comoving} specific intensity of a blackbody. The primes emphasize that both thermodynamic variables ($T'$) and emission angles ($\sigma'$) are defined in the instantaneous rest frame of the surface element. Our task below is to relate these {comoving} frame differentials to the ones measured by a distant observer.

We now calculate each term in the observer's frame. According to Doppler effect and gravatitional redshift,
\begin{equation}
E = \delta \sqrt{1-u} E' .
\end{equation}
The observed photon energy is boosted by the special-relativistic Doppler factor $\delta$ and reduced by the gravitational redshift factor $\sqrt{1-u}$, with $u\equiv R_{\rm S}/R$. 

According to number conservation, each emit photon should arrive the observer,
\begin{equation}
dN = dN' .
\end{equation}
According to Equation~(\ref{eq:dt_emit_receive}),
\begin{equation}
dt' = \frac{\sqrt{1-u}}{1 - \beta \cos\xi} dt = \sqrt{1-u} \delta \gamma dt .
\end{equation} 
The {comoving} time element $dt'$ relates to the observer frame time interval $dt$ by the same factors derived in Appendix~\ref{AppendixB:time_dilation}: gravitational time dilation (numerator) and kinematic arrival-time compression/expansion (denominator).

According to Equation~(\ref{eq:dOmega}),
\begin{equation}
\begin{split}
d\Omega' &= \delta^2 d\Omega = \delta^2 d\cos\alpha d\lambda \\
&= \delta^2 \frac{d\cos\alpha}{d\cos\psi} d\cos\psi d\lambda = \delta^2 \frac{d\cos\alpha}{d\cos\psi} \frac{dA}{D^2} ,
\end{split}
\end{equation}
where $\lambda$ is the azimuthal angle around the outward radial vector, $dA$ is the differential area of the detector, and $D$ is the distance from star to the observer.

The first equality is the standard formula $d\Omega'=\delta^2 d\Omega$ from Appendix~\ref{AppendixA:doppler}. We then (i) resolve $d\Omega$ into polar and azimuthal differentials around radial vector, (ii) introduce the gravitational light-bending Jacobian $d\cos\alpha/d\cos\psi$ that maps the apparent angle $\psi$ at infinity to local emission angle $\alpha$ of spherical surface, and (iii) relate $d\cos\psi d\lambda$ to detector's collecting area $dA$ via $d\cos\psi d\lambda=dA/D^2$.

Combine all these together, we have
\begin{equation}
\frac{EdN}{dtdAdE} = \sqrt{1-u} I_{bb} \delta^3 \gamma \cos\sigma' \frac{d\cos\alpha}{d\cos\psi} \frac{dS'}{D^2} .
\end{equation}
This is the energy flux per unit detector area and per unit energy. The prefactors have transparent origins: $\delta$ (from $E$), $\delta^2$ (from $d\Omega'$), and $\gamma$ (from the $dt'$ mapping) combine into $\delta^3\gamma$, $\sqrt{1-u}$ accounts for gravitational redshift, the cosine factor uses the {comoving} local emission angle $\sigma'$ , and the Jacobian encodes Schwarzschild gravitational lensing. The $dS'/D^2$ factor represent the differential area of the emitting patch and the stellar distance information.

In practice, we may substitute $I_{bb}$ with a more general intensity function $I(E',\sigma')$ to account for non-blackbody and anisotropic atmosphere beaming, and we want the number flux rather than energy flux, so we divide both side by $E$ and get
\begin{equation}
\frac{dN}{dtdAdE} = \sqrt{1-u} \delta^3 \gamma \cos\sigma' \frac{d\cos\alpha}{d\cos\psi} \frac{I(E',\sigma')}{E} \frac{dS'}{D^2} .
\end{equation}

\end{appendix}

\end{document}